\documentclass[aps,groupedaddress,longbibliography]{revtex4-1}

\usepackage{bbm}
\usepackage{color}
\usepackage{graphicx}
\usepackage{amssymb,latexsym}
\usepackage{tensor}
\usepackage{slantsc}
\usepackage{slashed}

\usepackage{mathtools}
\usepackage[normalem]{ulem}

\usepackage{marginnote}

\newcommand{\nored}[1]{{ #1}}

\newcommand{\boundaryset}{{\nored{\partial E(R)}}}

\newcommand{\Ome}[1]{\tensor{\overline{\Omega}}{#1}}

\newcommand{\KKN}{{{K}}}
\newcommand{\pKKN}{+\KKN}

\newcommand{\betaR}{{\nored{\beta_R}}}

\newcommand{\HboundX }{{H_b \left(X,\mcS \right)}}
\newcommand{\Hb  }{{H_b}}

\newcommand{\alphaz}{{\mathring \alpha}}
\newcommand{\betaz}{{\mathring \beta}}

\newcommand{\Sext}{\hypext}

 \newcommand{\nodiffGamma}{{\nored{\Gamma}}}
 \newcommand{\nodiffR}[1]{{\nored{\tensor{R}{#1}}}}
\newcommand{\nodiffKKR}[1]{\tensor{\nored{\mathbf{R}}}{#1}} 
\newcommand{\nodiffsptR}[1]{\nodiffR{#1}} 
\newcommand{\nodiffspR}[1]{\nored{\tensor{\mathcal{R}}{#1}}} 
\newcommand{\nodiffspKKR}[1]{\nored{\tensor{\textsf{R}}{#1}}} 

 \newcommand{\diffGamma}{{\nored{\delta \Gamma}}}
 \newcommand{\diffR}[1]{{\nored{\delta \tensor{R}{#1}}}}

\newcommand{\diffKKR}[1]{\delta \tensor{\nored{\mathbf{R}}}{#1}} 

 \newcommand{\dScoordinate}{{\nored{d\Sigma}}}

\newcommand{\Bgamma}{\backGamma{}}

\newcommand{\xm}{{ \tilde x}}
\newcommand{\xp}{{ \overline  x}}

 \newcommand{\fr}[1]{\nored{\hat{#1}}}
\newcommand{\Xzeroframe}{\nored{X^{\fr 0}}}

\newcommand{\Witten}{{\mycal W}}

\newcommand{\mcU}{{\mycal U}}
\newcommand{\ga}{\gamma}
\newcommand{\mtd}{\slashed{D}}

\newcommand{\NN}{{}^{\KKN}\!\!\!{\mycal N}}

\newcommand{\redtwo}{2}
\newcommand{\redsix}{6}

\newcommand{\spADM}{\nored{\textrm{ADM}}}
\newcommand{\spaceS}{\nored{\sigma}}

\newcommand{\rnu}{\nored{\nu}}

\newcommand{\nc}{\nored{\mathrm{n.c.}}}

\DeclareFontFamily{OT1}{rsfs}{}
\DeclareFontShape{OT1}{rsfs}{m}{n}{ <-7> rsfs5 <7-10> rsfs7 <10-> rsfs10}{}
\DeclareMathAlphabet{\mycal}{OT1}{rsfs}{m}{n}

\newcommand{\mcL}{{\mycal L}}

\newcommand{\mcS}{{\mycal S}}
\newcommand{\hyp}{\mcS}
\newcommand{\hypext}{\mcS_\ext}
\newcommand{\ext}{{\mathrm{ext}}}

{\catcode `\@=11 \global\let\AddToReset=\@addtoreset}
\AddToReset{equation}{section}

\newcounter{mnotecount}[section]

\newcommand{\T}{\mathbb{T}}

\newcommand{\back}[1]{\overline{#1}} 
\newcommand{\backg}{{{\overline{g}}}} 
\newcommand{\backGamma}{{{\overline{\Gamma}}}} 
\newcommand{\bcov}[1]{\back{\nabla}{#1}} 

\newcommand{\bKKR}[1]{\back{\KKRic}\tensor{\vphantom{R}}{#1}}
\newcommand{\dKKR}[1]{\nored{\tensor{\mathbf{R}}{#1}-\bKKR{#1}} }
\newcommand{\dsptR}[1]{\tensor{R}{#1}-\tensor{\overline {R}}{#1}} 
\newcommand{\dspR}[1]{\tensor{\mathcal{R}}{#1}-\tensor{\overline{\mathcal{R}}}{#1}} 
\newcommand{\dspKKR}[1]{\nored{\tensor{\textsf{R}}{#1}-\tensor{\overline{\textsf{R}}}{#1}}} 

\newcommand{\bnoKKW}[1]{\overline{{W}}\tensor{\vphantom{W}}{#1}}
\newcommand{\bnoKKR}[1]{\back{\noKKRic}\tensor{\vphantom{R}}{#1}}
\newcommand{\noKKW}[1]{\tensor{{{W}}}{#1}} 
\newcommand{\noKKR}[1]{\tensor{{{R}}}{#1}} 
\newcommand{\noKKRic}{{{R}}} 

\newcommand{\KKR}[1]{\tensor{\nored{\mathbf{R}}}{#1}} 
\newcommand{\KKRic}{\nored{\mathbf{R}}} 

\newcommand{\spKKR}[1]{\nored{\tensor{\textsf{R}}{#1}}} 

\newcommand{\eel}[1]{\label{#1}\end{equation}}
\newcommand{\eeal}[1]{\label{#1}\end{eqnarray}}

\newcommand{\mcK}{{\mycal K}}

\newcommand{\bel}[1]{\begin{equation}\label{#1}}
\newcommand{\bea}{\begin{eqnarray}}
\newcommand{\bean}{\begin{eqnarray}\nonumber}
\newcommand{\beal}[1]{\begin{eqnarray}\label{#1}}
\newcommand{\eea}{\end{eqnarray}}

\newcommand{\nn}{\nonumber}
\newcommand{\Eq}[1]{Equation~\eq{#1}}

\newcommand{\Eqs}[2]{Equations~\eq{#1}-\eq{#2}}

\newcommand{\be}{\begin{equation}}
\newcommand{\eeq}{\end{equation}}
\newcommand{\ee}{\end{equation}}
\newcommand{\beqa}{\begin{eqnarray}}
\newcommand{\eeqa}{\end{eqnarray}}
\newcommand{\beqan}{\begin{eqnarray*}}
\newcommand{\eeqan}{\end{eqnarray*}}
\newcommand{\ba}{\begin{array}}
\newcommand{\ea}{\end{array}}

\newcommand{\const}{\mbox{\rm const}} 

\newcommand{\mnote}[1]
{\protect{\stepcounter{mnotecount}}$^{\mbox{\footnotesize
$
\bullet$\themnotecount}}$ \marginpar{
\raggedright\tiny\em
$\!\!\!\!\!\!\,\bullet$\themnotecount: #1} }

\newcommand{\warn}[1]
{\protect{\stepcounter{mnotecount}}$^{\mbox{\footnotesize
$
\bullet$\themnotecount}}$ \marginpar{
\raggedright\tiny\em
$\!\!\!\!\!\!\,\bullet$\themnotecount: {\bf Warning:} #1} }

\newcommand{\R}{\mathbb R}

\newcommand{\eq}[1]{(\ref{#1})}

\newcommand{\beaa}{\begin{eqnarray*}}
\newcommand{\eeaa}{\end{eqnarray*}}

\def\ben{\begin{equation}}
\def\een{\end{equation}}
\def\bena{\begin{eqnarray}}
\def\eena{\end{eqnarray}}

\def\f(#1/#2){\frac{#1}{#2}}
\def\Frac(#1/#2){\left(\frac{#1}{#2}\right)}
\def\chris(#1-#2-#3){{\mit \Gamma}^{#1}{}_{{#2}{#3}} }
\def\tilchris(#1-#2-#3){\tilde{{\mit \Gamma}}^{#1}{}_{{#2}{#3}}}
\def\hatchris(#1-#2-#3){\hat{{\mit \Gamma}}^{#1}{}_{{#2}{#3}}}

\DeclareFontFamily{OT1}{rsfs}{}
\DeclareFontShape{OT1}{rsfs}{m}{n}{ <-7> rsfs5 <7-10> rsfs7 <10-> rsfs10}{}

\begin{document}

\title{Energy in higher-dimensional spacetimes\footnote{Preprint UWThPh-2017-24}}

\author{Hamed Barzegar}
\email[]{a1326719@unet.univie.ac.at}
\affiliation{Faculty of Physics and Erwin Schr\"odinger Institute,
University of Vienna}
\thanks{Supported in part by  the Austrian Science Fund (FWF) under {projects P23719-N16 and P29517-N27, } and by the Polish National Center of Science (NCN) under grant 2016/21/B/ST1/00940. We are also grateful to the Erwin Schr\"odinger Institute for Mathematics and Physics, University of Vienna, for hospitality and support during part of work on this paper.}

\author{Piotr T.\ Chru\'{s}ciel}
\email[]{piotr.Chru\'sciel@univie.ac.at}
\homepage[]{http://homepage.univie.ac.at/piotr.Chru\'sciel}
\affiliation{Faculty of Physics and Erwin Schr\"odinger Institute,
University of Vienna}

\author{Michael H\"orzinger}
\email[]{mi.hoerz@gmail.com}
\affiliation{Faculty of Physics and Erwin Schr\"odinger Institute,
University of Vienna}

\date{\today}

\begin{abstract}
We derive  expressions for the total Hamiltonian energy of gravitating systems in higher dimensional theories in terms of the Riemann tensor, allowing a cosmological constant $\Lambda \in \R$. Our analysis covers asymptotically anti-de Sitter spacetimes, asymptotically flat spacetimes, as well as  Kaluza-Klein asymptotically flat spacetimes. We show that the Komar mass equals the ADM mass in stationary asymptotically flat space-times in all dimensions, generalising the four-dimensional result of Beig, and that this is not true anymore with Kaluza-Klein asymptotics. We show that the Hamiltonian mass does not necessarily coincide with the ADM mass in Kaluza-Klein asymptotically flat space-times, and that the Witten positivity argument provides a lower bound for the Hamiltonian mass, and not for the ADM mass, in terms of the electric charge. We illustrate our results  on the five-dimensional Rasheed metrics, which we study in some detail, pointing out restrictions that arise from the requirement of regularity, seemingly unnoticed so far in the literature.
\end{abstract}


\maketitle

\tableofcontents

\section{Introduction}

A key notion in any physical theory is that of total energy, momentum, and similar global charges. The corresponding definitions, and their properties, depend  very much upon the asymptotic conditions satisfied by the fields. There are various possibilities here, dictated by the physical problem at hand. For instance, the vanishing and the sign of the cosmological constant play a crucial role. Next, one may find it convenient to use direct coordinate methods~\cite{ADMGoldie,AbbottDeser,AshM}, or conformal methods~\cite{TanabeTanahashiShiromizu,AshtekarHansen}, or else~\cite{AshtekarRomano}, to define the asymptotic conditions and the objects at hand. Finally, one may want to use definitions arising from Hamiltonian techniques~\cite{AshBR,BeigOMurchadha87}, or appeal to the Noether theorem~\cite{WaldZoupas}, or use ad-hoc conserved currents~\cite{JJCYKADS,JJCYKCQG,JJCYK,KastorTraschenYano,LazkozSenovillaVera}. See also \cite{Trautman:Witten} for an excellent review of early work on the subject.

A natural class of asymptotic  conditions arises when considering isolated systems in Kaluza-Klein-type theories, see Section~\ref{s14II17.1} below. Much to our surprise, no systematic study of the notion of energy in this context appears to exist in the literature, and one of the aims of this work is to fill this gap. For this, we derive new expressions for the total Hamiltonian energy in higher-dimensions in terms of the Riemann tensor, in asymptotically flat, or asymptotically Kaluza-Klein, or asymptotically anti-de Sitter space-times. Our definitions arise from a Hamiltonian analysis of the fields and invoke direct coordinate- or tetrad-based asymptotic conditions. We relate these integrals to Komar-type integrals. We use Witten's argument to derive global inequalities between the Hamiltonian energy-momentum and the Kaluza-Klein charges.
We test our energy expressions on the Rasheed family of five dimensional vacuum metrics, clarifying furthermore some aspects of the  global structure of these solutions.

This paper is organised as follows: In Section~\ref{s14II17.1} we make precise our notion of Kaluza-Klein asymptotic flatness. At the beginning of Section~\ref{s14II17.2} we review the definition of energy within the  Hamiltonian framework of \cite{KijowskiTulczyjew,ChAIHP}. In Section~\ref{ss17IV17.1} we apply the framework to space-times which are asymptotically flat in a Kaluza-Klein sense. In Section~\ref{ss22IV17.1} we derive general formulae which apply for a large class of asymptotic conditions. In Section~\ref{s14II17.3} we show how to rewrite the formulae derived so far in terms of the curvature tensor. This is done in Section~\ref{ss21.IV.17} for KK-asymptotically flat solutions, and in Section~\ref{ss20IV17.2} for general backgrounds. The formulae are then specialised in Section~\ref{ss23VI17.2} to asymptotically anti-de Sitter solutions, and in Section~\ref{ss23VI17.1} to a class of Kaluza-Klein solutions with vanishing cosmological constant which are not KK-asymptotically flat.
In Section~\ref{ss5IV17.1} we rewrite some of our Riemann-integral energy expressions in terms of a space-and-time decomposition of the metric. In Section~\ref{22II17} we show how to establish Komar-type expressions for energy in space-times with Killing vectors. In Section~\ref{s14II17.5} we show how a Witten-type positivity argument applies to obtain global inequalities for KK-asymptotically flat metrics. Appendix~\ref{s14II17.6} is devoted to a study of the geometry of Rasheed's Kaluza-Klein black holes, which provide a non-trivial family of examples for which our energy expressions can be explicitly calculated.

 \section{Kaluza-Klein asymptotics}
  \label{s14II17.1}

 The starting point for our notion of Kaluza-Klein asymptotics are  initial data surfaces in an $(n\pKKN +1)$-dimensional space-time containing  asymptotic ends of the form
\begin{equation}
 \label{2II17.1}
 {\mycal S}_\ext :=\left(
  \mathbb{R}^n \setminus B(0,R)
   \right) \times \underbrace{S^1\times \cdots \times S^1}_{\textrm{$\KKN $ factors}} =: \left( \mathbb{R}^n \setminus B(0,R)
   \right) \times  \mathbb{T}^\KKN
  \,
\end{equation}
where $S^1$ is the unit circle.   We  will say that the metric is KK-asymptotically flat if  $g$ has the following asymptotic form along $\hypext\equiv \{x^0=0\}$:
\begin{equation}
 \label{2II17.2}
 g= \underbrace{\eta_{ab}dx^a dx^b + \delta_{AB} dx^A dx^B}_{=:\eta_{\mu\nu}dx^\mu dx^\nu} + o (r^{-\alpha})
 \,,
  \quad
  \partial_\mu g_{\nu\rho} =  o (r^{-\alpha-1})
 \,,
\end{equation}
where greek indices run from $0$ to $n\pKKN $, upper case latin indices  from the beginning of the alphabet  run from $n+1$ to $n\pKKN $,  lower case latin indices from the beginning of alphabet running from $0$ to $n $,  and lower case latin indices from the middle of alphabet running from $1$ to $n$.  Finally, upper case latin indices from the middle of the alphabet  run from $1$ to $n\pKKN $. Summarising:
\bel{17IV17.1}
 (x^\mu)\equiv (x^0,x^i,x^A)\equiv (x^a,x^A) \equiv (x^0,x^I)
  \,.
\ee
Last but not least,
$$
 r:=\sqrt{(x^1)^2+\ldots+(x^n)^2}
  \,.
$$
The exponent $\alpha$ will be chosen to be the optimal-one for the purpose of a well posed definition of total energy, namely
\bel{14II17.1}
 \alpha = \frac{n-2}2
 \,,
\ee
where, as in \eq{2II17.1}, $n$ is the space-dimension \emph{without counting the Kaluza-Klein directions}.

 In Kaluza-Klein theories it is often assumed that the vector fields $\partial_A$ are Killing vectors, but we will not make this hypothesis unless explicitly indicated otherwise.

 \section{Hamiltonian charges}
  \label{s14II17.2}

In this section we adapt the Hamiltonian analysis of~\cite{ChAIHP} (based on \cite{KijowskiTulczyjew}, compare~\cite{CJK}) to the asymptotically KK setting, providing also convenient alternative expressions for the formulae for the Hamiltonians derived there. We use a background metric $\backg_{\mu\nu}$,
which is assumed to be asymptotically KK as defined in Section~\ref{s14II17.1}, to determine the asymptotic conditions. The metric $\backg_{\mu\nu}$ should be thought of as being the metric $
\eta_{\mu\nu}$ of Section~\ref{s14II17.1} at large distances, but it might be convenient in some situations to use coordinate systems where $\backg_{\mu\nu}$ does not take an explicitly flat form.

Every such  metric $\backg_{\mu\nu}$ determines a family of metrics $g_{\mu\nu}$ which asymptote to it in the sense of \eq{2II17.2}. We will denote by $\backGamma{^\alpha_{\beta\gamma}}$ the Christoffel symbols of the Levi-Civita connection of $\backg_{\mu\nu}$.

Given a vector field $X$, the calculations in~\cite{ChAIHP} show that the flow of $X$ in the space-time obtained by evolving the initial data on $\hyp$ is Hamiltonian with respect to a suitable symplectic structure, with a Hamiltonian $H(X,\mcS)$ which, in vacuum,
 is  given by the formula
\begin{equation}\label{a.1}
H(X,\mcS )=\int_\mcS  \left(p^\mu_{\alpha\beta}\mathcal{L}_X\mathfrak{g}^{\alpha\beta}-X^\mu L\right)\dScoordinate_\mu\,,
\end{equation}
where
\bean
L &\coloneqq &
 \mathfrak{g}^{\mu\nu}\left[\left(\tensor{\Gamma}{^\alpha_{\sigma \mu}}-\tensor{\backGamma}{^\alpha_{\sigma \mu}}\right)\left(\tensor{\Gamma}{^\sigma_{\alpha \nu}}-\tensor{\backGamma}{^\sigma_{\alpha \nu}}\right)-\left(\tensor{\Gamma}{^\alpha_{\mu\nu}}-
  \tensor{\backGamma}{^\alpha_{\mu\nu}}\right)\left(\tensor{\Gamma}{^\sigma_{\alpha\sigma}}
   -\tensor{\backGamma}{^\sigma_{\alpha\sigma}}\right)+\bKKR{_{\mu\nu}}
   {- \frac{2}{d+\KKN}\Lambda g_{\mu\nu}}
    \right]
\\
 &&
 - \frac{1}{16\pi}\sqrt{-\det \backg}\, \backg^{\mu\nu}\big(\bKKR{_{\mu\nu}}
   {- \frac{2}{d+\KKN}\Lambda \backg_{\mu\nu}}
   )
     \,,
\eeal{2V17.1}
with $\bKKR{_{\mu\nu}}$ being the Ricci tensor of the background metric $\backg_{\mu\nu}$, $\Lambda$ the cosmological constant, $d$ the dimension of the physical space-time, $\KKN$ is the number of Kaluza-Klein dimensions (possibly zero),
  and
\begin{equation}
 \label{30IV17.2}
\mathfrak{g}^{\mu\nu}\coloneqq \frac{1}{16\pi}\sqrt{-\det g}\, g^{\mu\nu}
 \,,
 \quad
 p^\lambda_{\mu\nu}
  :=
   \frac{\partial L}{\partial\tensor{\mathfrak{g}}{^\mu^\nu_{,\lambda}}}
   =
  \left( {\Bgamma}^{\lambda}_{\mu\nu} -
  {\delta}^{\lambda}_{(\mu} {\Bgamma}^{\kappa}_{\nu ) \kappa} \right) -\left({\Gamma}^{\lambda}_{\mu\nu} - {\delta}^{\lambda}_{(\mu}
{\Gamma}^{\kappa}_{\nu ) \kappa}\right)
 \,.
\end{equation}
%
%
%
Finally, the volume forms $\dScoordinate_{\alpha }$ and $\dScoordinate_{\alpha\beta}$ are defined as
\bel{22IV17.2}
 \dScoordinate_{\alpha} = \partial_\alpha \rfloor (dx^0\wedge \cdots \wedge dx^{n\pKKN })
  \,,
  \quad
 \dScoordinate_{\alpha \beta } = \partial_\beta  \rfloor \dScoordinate_{\alpha}
 \,,
\ee
where $\rfloor$ denotes the contraction:  for any vector field $X$ and skew-form $\alpha$ we have $X\rfloor \alpha(\cdot,\ldots) := \alpha(X,\ldots)$.

We note that the last two, $g$-independent, ``renormalisation'' terms in \eq{2V17.1} have been added for convergence of the  integrals at hand.

We will write $\det g \equiv \det (g_{\mu\nu})$ for the determinant of the full metric tensor, writing explicitly $\det (g_{IJ})$ for the determinant of the metric $g_{IJ}dx^Idx^J$ induced on the level sets of $x^0$, etc., when need arises.

We emphasise that the formal considerations in~\cite{ChAIHP} are quite general, applying regardless of the asymptotic conditions and of dimensions. However, the question of convergence and well posedness of the resulting formulae appears to require a case-by-case analysis, once a set of asymptotic conditions has been imposed.

If $X$ is a Killing vector field of $\backg_{\mu\nu}$  and if the Einstein equations  with sources  and with a cosmological constant $\Lambda$ are satisfied,
\bel{3V17.1}
 R_{\mu\nu} - \frac 12 R g_{\mu\nu} + \Lambda g_{\mu\nu} = 8 \pi T_{\mu\nu}
 \,,
\ee
the integrand \eqref{a.1} can be rewritten as the divergence of a ``Freud-type superpotential'', up to source and renormalisation terms:
\begin{equation}
 \label{30IV17.11}
H^\mu\equiv p^\mu_{\alpha\beta}\mathcal{L}_X \mathfrak{g}^{\alpha\beta}
 -X^\mu
   L
    =\partial_\alpha \mathbb{U}^{\mu\alpha}
 -
   \sqrt{-\det g }\,
 T^\mu{}_\alpha X^\alpha
 +
   \frac{1}{16\pi}
   \sqrt{-\det \backg }\,
   \backg^{\alpha \beta}\big(\bKKR{_{\alpha \beta}}
   {- \frac{2}{d+\KKN}\Lambda \backg_{\alpha \beta}})X^\mu
     \,,
\end{equation}
with
\bea\label{a.7}
&
 \mathbb{U}^{\nu\lambda}
 =
  \tensor{\mathbb{U}}{^\nu^\lambda_\beta}X^\beta
  -
  \frac{1}{8\pi}\sqrt{|\det g|} \, g^{\alpha[\nu}\delta^{\lambda]}_\beta
  {\overline \nabla_{  \alpha}}\tensor{X}{^\beta}
  \,,
  &
\\
&
\tensor{\mathbb{U}}{^\nu^\lambda_\beta}=\frac{2 |\det \backg |}{16\pi\sqrt{|\det g|}}g_{\beta\gamma}
 {{\overline \nabla _ \kappa}}\left(e^2 g^{\gamma[\lambda}g^{\nu]\kappa}\right)
\,,
&
\eea
where $\overline \nabla$
 denotes the covariant derivative of the background metric $\backg_{\mu\nu}$ and
\begin{equation}
e^2\equiv \frac{\det g}{\det \backg }\,.
\end{equation}
 In vacuum  this leads to the formula
\begin{equation}
 \label{14II17.2}
H \left(X,\mcS \right) =
\HboundX :=\frac{1}{2}\int_{\partial\mcS }
 \big( \mathbb{U}^{\nu\lambda} - \mathbb{U}^{\nu\lambda}  \big|_{g=\backg}\big)\dScoordinate_{\nu\lambda}
 \,,
\end{equation}
where the subscript ``$b$'' on $\Hb$ stands for ``boundary''.
For vector fields $X$ which are not necessarily Killing vector fields of the background, the Hamiltonian might have some supplementary volume terms, cf.~\cite{CJK,CJKKerrdS}.
In non-vacuum Lagrangian diffeomorphism-invariant field theories, this formula for the total Hamiltonian of the coupled system of fields remains true after adding to $H^\mu$ a contribution from the matter fields; cf., e.g.,~\cite{KijowskiTulczyjew,KijowskiGRG,CJKKerrdS}.

\subsection{Kaluza-Klein asymptotics}
 \label{ss17IV17.1}

For Kaluza-Klein asymptotically flat field configurations we have
\begin{equation}
 \label{1II17.1}
 g_{\mu\nu}=\eta_{\mu\nu}+o\left(r^{-\alpha}\right)
 \,,
 \quad
 \partial_\sigma g_{\mu\nu}= o\left(r^{-\alpha-1 }\right)
 \,,
 \quad
 \backg_{\mu\nu}=\eta_{\mu\nu}+o\left(r^{-\alpha}\right)
 \,,
 \quad
 \partial_\sigma \backg_{\mu\nu}= o\left(r^{-\alpha-1 }\right)
 \,.
\end{equation}
In particular this implies
$$
 \tensor{\backGamma}{^\alpha_{\beta \gamma}} = o(r^{-\alpha-1})
 \,.
$$

Let us,  first, assume that $X$ is   $\backg$-covariantly constant   (hence also a Killing vector of the background metric $\backg_{\mu\nu}$). One then checks that in the coordinates of \eq{1II17.1} the vector field $X$ has to be of the form
\bel{14II17.5}
 X^\mu =X^\mu_\infty + o(r^{-\alpha})
 \,,
 \quad
 \partial_\nu X^\mu_\infty = 0
  \,.
\ee

As $\Lambda=0$ in the current case, convergence of the boundary integrals in vacuum will be guaranteed if one assumes, e.g.,
\bel{30IV17.1}
 \sum_{\mu \alpha\beta}\int_{\hyp \cap \{r\ge R\}} |\partial_\mu g_{\alpha\beta}|^2   d^{n\pKKN }x <\infty
  \,.
\ee
This follows immediately from Stokes' theorem together with \eq{a.1}-\eqref{30IV17.2} and \eqref{30IV17.11}, keeping in mind that $\Lambda=0=\bKKR{_{\mu\nu}}$ in the current context.

We note that \eq{30IV17.1} will hold if \eq{14II17.1} is replaced by $\alpha > (n-2)/2$, which provides a sufficient but not a necessary  condition.

While we are mostly interested in vacuum solutions, the analysis below applies to non-vacuum ones, provided that one also has
\bel{30IV17.4}
 T_{\mu\nu} = o (r^{-n })
  \
  \mbox{and}
  \
 \sum_{  \alpha\beta} \int_{\hyp \cap \{r\ge R\} } |T_{\alpha\beta}|  d^{n\pKKN }x <\infty
 \,.
\ee
\Eqs{30IV17.1}{30IV17.4} will be assumed in the calculations that follow.

Since the last term in  \eqref{a.7} drops out when $\overline \nabla_\beta X_{\alpha }=0$, we obtain
\begin{eqnarray}
 \nn
 \mathbb{U}^{\nu\lambda}
 &=&
  \tensor{\mathbb{U}}{^\nu^\lambda_\beta} X^\beta \nonumber
\\
\nn
 &=&
  -\frac{1}{16\pi}
   \left(
    1+o(r^{-\alpha})
   \right)
   \left(
    \eta_{\beta\gamma}+o(r^{-\alpha})
   \right) X^\beta
  \Big[ \left(
    \eta^{\gamma\nu}\eta^{\lambda\kappa}\eta_{\rho\sigma}
    -
    \eta^{\gamma\lambda}\eta^{\nu\kappa}\eta_{\rho\sigma}
    \right) \tensor{g}{^{\rho\sigma}_{,\kappa}}
\\
\nn
 &&
  +
   \tensor{g}{^{\gamma\lambda}_{,\kappa}} \eta^{\nu\kappa}
   -
   \tensor{g}{^{\gamma\nu}_{,\kappa}} \eta^{\lambda\kappa}
   +
   \eta^{\gamma\lambda}\tensor{g}{^{\nu\kappa}_{,\kappa}}
   -
   \eta^{\gamma\nu}\tensor{g}{^{\lambda\kappa}_{,\kappa}}+o\left(r^{-2\alpha-1}\right)
   \Big]
\\
\nn
 &=&
   -\frac{1}{16\pi}
   \Big[ \left(
   \eta^{\lambda\kappa}X^\nu
   -
   \eta^{\nu\kappa}X^\lambda\right)\eta_{\rho\sigma}\tensor{g}{^{\rho\sigma}_{,\kappa}}
   +
   \eta^{\nu\kappa}\eta_{\beta\gamma}\tensor{g}{^{\gamma\lambda}_{,\kappa}}X^\beta
   -
   \eta^{\lambda\kappa}\eta_{\beta\gamma}\tensor{g}{^{\gamma\nu}_{,\kappa}}X^\beta
\\
\nn
 &&
 +
 \tensor{g}{^{\nu\kappa}_{,\kappa}}X^\lambda
 -
 \tensor{g}{^{\lambda\kappa}_{,\kappa}}X^\nu
 \Big]
  +
  o\left(r^{-2\alpha-1}\right)
\\
\nn
 &=&
   -\frac{1}{16\pi}\eta^{\delta\kappa}\eta_{\beta\gamma}
   \tensor{g}{^{\gamma\tau}_{,\kappa}}X^\xi
   \Big(
    \delta^{\lambda}_{\delta}\delta^{\nu}_{\xi}\delta^{\beta}_{\tau}
    -
    \delta^{\lambda}_{\xi}\delta^{\nu}_{\delta}\delta^{\beta}_{\tau}
\\
\nn
 &&
  +
  \delta^{\lambda}_{\tau}\delta^{\nu}_{\delta}\delta^{\beta}_{\xi}
  -
  \delta^{\lambda}_{\delta}\delta^{\nu}_{\tau}\delta^{\beta}_{\xi}
  +
  \delta^{\lambda}_{\xi}\delta^{\nu}_{\tau}\delta^{\beta}_{\delta}
  -
  \delta^{\lambda}_{\tau}\delta^{\nu}_{\xi}\delta^{\beta}_{\delta}\Big)
  +
  o\left(r^{-2\alpha-1}\right)
 \\
 &=&
 \frac{3}{8\pi}\eta^{\delta\kappa}\eta_{\beta\gamma}
  \tensor{g}{^{\gamma\tau}_{,\kappa}}X^\xi\,\delta^{\nu\lambda\beta}_{\tau\delta\xi}
  +
  o\left(r^{-2\alpha-1}\right)
  \,.
\end{eqnarray}
Plugging the result into \eqref{a.7} and renaming indices, in the limit $r\rightarrow\infty$, we obtain the following form of \eq{14II17.2}, which will be seen to be convenient in our further considerations:
\begin{equation}\label{a.12}
 \HboundX
  =
   \frac{3}{16\pi}\lim_{R\rightarrow\infty}\int_{S(R)\times \mathbb{T^\KKN }}
    \delta^{\alpha\beta\gamma}_{\lambda\mu\nu} X^\nu\eta^{\lambda\rho}\eta_{\gamma\sigma}\partial_\rho g^{\sigma\mu} dS_{\alpha\beta}\,,
\end{equation}
where $S(R)$ denotes a sphere of radius $R$ in the $\R^n$ factor of $\hyp_{\ext}$,
and
\begin{equation}
\delta^{\alpha\beta\gamma}_{\lambda\mu\nu}\coloneqq\delta^\alpha_{[\lambda}\delta^\beta_\mu\delta^\gamma_{\nu ]}\,.
\end{equation}

We see from \eq{14II17.5} that $\HboundX $ can be written as
\begin{equation}
\HboundX \eqqcolon p_\mu X^\mu_{\infty}\,.
\end{equation}
When $\KKN =0$ the coefficients $p_\mu$ are called the ADM four-momentum of $\mcS $ \cite{ADMGoldie}.

 If $X=\partial_0$  we find a formula somewhat resembling the usual one:
\newcommand{\boundarymeasure}{\nored{d^{n\pKKN -1}\mu}}
\begin{eqnarray}
 p_0 &:=&  \Hb \left(\partial_t,\mcS \right)
  =
   \frac{1}{16\pi}\lim_{R\rightarrow\infty}\int_{S(R)} \int_{\mathbb{T}^\KKN }
   \sum_{I=1}^{n\pKKN }( \partial_I g_{i I } - \partial_i g_{I I}) \frac{x^i}R  \, \boundarymeasure  
 \nn
\\
 &
  =
   &
   |\T^\KKN | \, p_{0,\spADM} +
   \frac{1}{16\pi}\lim_{R\rightarrow\infty}\int_{S(R)} \int_{\mathbb{T}^\KKN }
   \sum_{A=n+1}^{n\pKKN }( \partial_A g_{i A } - \partial_i g_{AA}) \frac{x^i}R  \, \boundarymeasure  
    \,.
    \label{a.12++}
\end{eqnarray}
Here $\boundarymeasure$ is the measure induced on $S(R)\times \T^\KKN $ by the flat metric, $|\T^\KKN |$ denotes the volume of $\T^\KKN $, and $p_{0,\spADM}$ is the usual  (total) ADM energy  of the physical-space metric $g_{ij}dx^i dx^j$. Perhaps not unexpectedly, the ADM energy $p_{0,\text{ADM}}$  does \emph{not} coincide with the Hamiltonian generating time-translations in general.

Next, when $X^0=0$, after using Stokes theorem in the following integral
\bel{24III17.31}
 \int_{S(R)\times \T^\KKN } \partial_J( g_{J0}\delta_I^{L} - g_{L0} \delta_I^J) \partial_{L} \rfloor (dx^1\wedge \cdots\wedge  \nored{dx^{\KKN+ n}} )= 0
 \,,
\ee
 we obtain the formula
%
\begin{equation}\label{24III17.21}
 p_I    := \Hb \left( \partial_I,\mcS \right)
  =
 \frac{1}{8 \pi R} \lim_{R\rightarrow\infty}\int_{S(R)} \int_{\mathbb{T}^\KKN }
    P_{Ii} x^i \, \boundarymeasure  
    \,.
\end{equation}
Here $P_{IJ}$ is the usual  canonical  ADM momentum
\bel{24III17.21+}
 P_{IJ} := g^{LM}k_{LM} g_{IJ} - k_{IJ}
 \,,
 \quad
 k_{IJ} :=  \frac 12 \mcL_T g_{IJ} =   \frac 12 (\partial_0 g_{IJ} - \partial_I g_{0J} - \partial_J g_{0I}) + o(r^{-2\alpha -1})
  \,,
\ee
while $\mcL_T$ denotes the Lie derivative in the direction of the unit-timelike future directed field $T$    of normals to the level sets of $ x^0 $.

As an example, we compute the above integrals for the Rasheed metrics, described in Appendix~\ref{s14II17.6},  with $P=0$:
\bel{31III17.12}
 p_0 = 2\pi  \nored{M
 }
  \,,
  \quad
   p_i = 0
  \,,
  \quad
   p_4 = 2 \pi  Q
 \,.
\ee
\Eq{31III17.12} includes  a $2\pi$ factor arising from a normalisation in which the Kaluza-Klein coordinate $x^4$ in the Rasheed solutions runs over a circle of length $2\pi$.

This should be compared with the ADM four-momentum $p_{\mu,\spADM}$ of the $n$-dimensional space metric $g_{ij}dx^i dx^j$, which reads
%
\bel{31III17.12+=a}
 p_{0,\spADM} = \nored{M-\frac{\Sigma}{\sqrt{3}}}
  \,,
  \quad
   p_{i,\spADM} = 0
 \,.
\ee

\subsection{General backgrounds}
 \label{ss22IV17.1}

As discussed in detail in Appendix~\ref{s24IX17.1}, the Rasheed solutions with $P\ne 0$ are not KK asymptotically flat in the sense set forth above. To cover this case we need to generalise the calculations so far to the case where the background metric is not flat, with an asymptotic region  $\hypext\subset \hyp$ diffeomorphic to
\bel{20IV17.4}
 \hypext \approx E(R_0)\,, \ \mbox{where} \ E(R):=\big(\R^n \setminus B(R)\big)\times \NN
 \,,
\ee
with some $\KKN $-dimensional compact manifold $\NN$, for some $R_0\ge 0$. We therefore have an associated global coordinate system $x^i$ on $
 \R^n \setminus B(R_0)$,   as well as the dilation vector field $Z=x^i\partial_i\equiv r \partial_r$ which will play a key role in some calculation below.

Somewhat more generally, in order to be able to include general ``Birmingham-Kottler-Schwarzschild anti-de Sitter''   metrics, we will consider ends $E(R)$ equipped with a radial function $r$ so that
\bel{20IV17.4+}
 \hypext \approx  E(R_0)\,, \ \mbox{with} \ E(R):=\{r\ge R\}\equiv [R,\infty)\times \mcK
 \,,
\ee
where $\mcK$ is a compact manifold. Here $r$ is a coordinate running along the $[R_0,\infty)$ factor of $\hypext$, and the dilation vector $Z$ is defined as $Z:=r\partial_r$.

For the usual $(n+1)$-dimensional Schwarzschild-anti-de Sitter metric the manifold  $\mcK$ will be an $(n-1)$-dimensional sphere, but it can be an arbitrary compact manifold admitting Einstein metrics in the case of metrics  \eq{24V17.21}-\eq{24V17.22} below.

Along $\hypext$
we are given two Lorentzian metrics $g$ and $\backg$, with $g$ asymptotic to the background $\backg$ in a sense which we make precise now. Denoting by $\back{\nabla}$ the Levi-Civita connection associated with $\backg$, we assume the existence of a $\backg$-orthonormal frame $\{\overline  e_\fr \mu\}$ defined along $\hypext$ such that, decorating frame-indices with hats,
\begin{equation}
 g_{\fr \mu\fr \nu}:= g(\overline  e_\fr \mu, \overline  e_\fr \nu)
  =
 \backg_{\fr \mu\fr \nu}+o \left( r^{-\alpha} \right)
 \,,
 \qquad
 \back{\nabla}_\fr\lambda g  _{\fr \mu\fr \nu} = o \left( r^{-\beta} \right)
 \,.
  \label{20VI17.1}
\end{equation}
It seems that the specific values of  $\alpha$ and $\beta$ as needed for our mass formulae can only be chosen after a case-by-case study of the background metric $\backg$; compare \eq{19IV17.1+}-\eq{2V18/2} below.

In what follows we will use the following convention: given two tensor fields $u$ and $v$, we will write
\bel{25V17.1}
 u=v + o(r^{-\alpha})
\ee
if the frame components of $u-v$, within the class of $\backg$-ON frames chosen, decay as $o(r^{-\alpha})$. If $\overline  e_\fr 0$ is orthogonal to $\Sext$ (which will often be assumed)  then, if we denote by $\backg_\hyp:=\backg_{IJ}dx^Idx^J$ the Riemannian metric induced by $\backg $ on $\hypext$,  and by $|\cdot|_{\backg_\hyp}$ the associated norm, we have e.g.
$$
 u_{\mu\nu}= o( r^{-\alpha})
 \quad
  \Longleftrightarrow
  \quad
 | u_{\fr 0 \fr 0} | + |u_{\fr 0 I}dx^I |_{\backg_\hyp} + |u_{I J}dx^I dx^J |_{\backg_\hyp} = o( r^{-\alpha})
 \,.
$$

Assuming again that $X$ is $\backg$-covariantly constant, the second term of \eqref{a.7} vanishes and for the first term we have the same expression as in the KK-asymptotically flat case, with the difference that instead of $\eta_{\mu\nu}$ we have $\backg_{\mu\nu}$ and instead of partial derivatives we have covariant derivatives of the background metric, i.e.,
\begin{eqnarray}
 \nn%
 \mathbb{U}^{  \nu \lambda}
 &=&
  \tensor{\mathbb{U}}{^{  \nu}^{ \lambda}_{ \xi}}X^ \xi
  \\
  &=&
  \left(
  \frac{3}{8\pi}
  \delta^{\nu\lambda\sigma}_{\tau\delta\xi}
  \backg^{\delta\kappa}\backg_{\sigma\gamma}
  X^\xi
  \back{\nabla}_\kappa \tensor{g}{^{\gamma\tau} }
  +
  o\left( |X| r^{- \alpha-\beta}\right)\right)
   \sqrt{|\det g |}
  \,,
   \label{2V17.5}
\end{eqnarray}
where
\bel{2V17.2}
 |X|^2:= \sum_{\mu} (X^{\fr \mu})^2
 \,.
\ee

In order to control the error terms appearing in \eq{2V17.5} we will assume that
\bean
 &&
   \mbox{$\alpha$ and $\beta$ are such that the subleading terms $
  o\left(|X| r^{-\alpha-\beta}\right)$ in \eq{2V17.5} give}
\\
 &&
    \mbox{vanishing contribution to the boundary integrals after passing to the limit.
     }
\eeal{19IV17.1+}

This will e.g.\ be the case for all Rasheed metrics when $\alpha= (n-2)/2$ as in \eq{14II17.1}, $\beta=\alpha+1$, with $X$ asymptotic to $\partial_\mu$ in coordinates as in \eq{31III17.11}.

Similarly \eq{19IV17.1+} will be satisfied for asymptotically anti-de Sitter metrics with
\bel{2V18/2}
 \alpha=\beta=n/2
 \,,
\ee
where $r$ is the area coordinate for the anti-de Sitter metric. Note that in this case we have $|X|= O(r)$.

Instead of \eq{a.12} we obtain now
\begin{equation}
 \label{30IV17.31}
  \HboundX
   =
   \frac{3}{16\pi}
   \lim_{R\rightarrow\infty}\int_{\nored{\partial E(R)}}
   \delta^{\alpha\beta\gamma}_{\lambda\mu\nu}
   X^\nu\backg^{\lambda\rho}\backg_{\gamma\sigma}\bcov{_\rho} g^{\sigma\mu}
   dS_{\alpha\beta}\,,
\end{equation}
where the two-forms $ dS_{\alpha\beta}$ in $d\pKKN  \equiv n+1\pKKN $ space-time dimensions take the form
\begin{equation}\label{2.1}
dS_{\alpha\beta}=\frac{1}{(n\pKKN -1)!}\tensor{\epsilon}{_\alpha_\beta_{\xi_1}_\cdots_{\xi_{n\pKKN -1}}}\, dx^{\xi_1}\wedge\cdots\wedge dx^{\xi_{n\pKKN -1}}
  \equiv  \partial_\beta \rfloor \partial_\alpha \rfloor \underbrace{\sqrt{|\det g|} dx^0\wedge \cdots \wedge dx^{n\pKKN }}_{=: d\mu_g}
  \,.
\end{equation}

We can now compute the Hamiltonian charges for this general case. We have
\begin{eqnarray}
 \nn
  \lefteqn{
   \frac{3}{16\pi}
    \delta^{\alpha\beta\gamma}_{\lambda\mu\nu}
    X^\nu \backg^{\lambda\rho}\backg_{\gamma\sigma}\bcov{_\rho} g^{\sigma\mu}
    dS_{\alpha\beta}
  }
  &&
  \\
  \nn
  &=&
   \frac{1}{16\pi}
    \left(
     \delta^{\alpha\beta}_{\lambda\mu} \delta{^\gamma_\nu}
     +
     \delta^{\alpha\beta}_{\mu\nu} \delta{^\gamma_\lambda}
     +
     \delta^{\alpha\beta}_{\nu\lambda} \delta{^\gamma_\mu}
    \right)
   X^\nu \backg^{\lambda\rho} \backg_{\gamma\sigma}\bcov{_\rho} g^{\sigma\mu}
    dS_{\alpha\beta}
    \\
    &=&
    \frac{1}{16\pi}
     \left(
       X^\gamma \backg^{\lambda\rho} \backg_{\gamma\sigma}
       \bcov{_\rho} g^{\sigma\mu} dS_{\lambda\mu}
       +
        X^\nu \bcov{_\sigma} g^{\sigma\mu} dS_{\mu\nu}
       +
       X^\nu \backg^{\lambda\rho} \backg_{\gamma\sigma}
       \bcov{_\rho} g^{\sigma\gamma} dS_{\nu\lambda}
     \right)
     \,.
\end{eqnarray}

To continue, it is best to use an $\backg$-orthonormal
 frame $\overline{e}_\fr i$ with $\overline{e}_{\fr 0}$ orthogonal to $\hyp$ and $\overline{e}_{\fr A}$ tangent to $\partial E(R)$. Then only the forms $dS_{\fr 0\fr i}$ give a non-vanishing contribution to the boundary integral.
In the calculations that follow we will write ``$\nc$''
 for the sum of those terms which do not contribute to the integral either because of the integration domain, or by Stokes theorem, or by passage to the limit.

If $X=\partial_0$, and assuming that
\bel{22IV17.1}
 \partial_0=\Xzeroframe\overline{e}_{\fr 0}
\ee
one finds, using frame indices throughout the calculation,
\begin{eqnarray}
 \nn
   \frac{3}{16\pi}
    \delta^{\alpha\beta\gamma}_{\lambda\mu\nu}
    X^\nu \backg^{\lambda\rho}\backg_{\gamma\sigma}\bcov{_\rho} g^{\sigma\mu}
    dS_{\alpha\beta}
 &=&
   \frac{3}{8\pi} \Xzeroframe
    \delta^{\fr 0\fr k\fr J}_{\fr \lambda\fr \mu\fr  0}
     \backg^{\fr \lambda\fr \rho}\backg_{\fr J\fr \rnu }\bcov{_{\fr \rho}}
    g^{\fr \rnu \fr \mu}
    dS_{\fr 0\fr k}
     \\
     &=&
      \frac{1}{16\pi} \Xzeroframe
       \left[
        \bcov{^\fr k}
             \left(
              \backg_{\fr J \fr L} g^{\fr J \fr L}
             \right)
          -
         \bcov{_{\fr J} } g^{\fr J \fr k   }
       \right]
       d\spaceS_{\fr k}
       +\nc
        \,,
\end{eqnarray}
where
\bel{19IV17.1}
 d\spaceS_{\fr i} := dS_{\fr 0\fr i}
  \equiv\overline{e}_{\fr i} \rfloor \overline{e}_{\fr 0} \rfloor ( \sqrt{|\det g|} dx^0\wedge \cdots \wedge dx^{n\pKKN })
  \nored{
  =\overline{e}_{\fr i} \rfloor  ( \sqrt{|\det g_{IJ}|} dx^1\wedge \cdots \wedge dx^{n\pKKN }) + \nc
  }
\ee
Hence, we obtain the following generalisation of the ADM energy:
\begin{equation}
 p_0
  =
  \Hb \left(\partial_t,\mcS \right)
  =
   \frac{1}{16\pi}\lim_{R\rightarrow\infty}\int_{\nored{\partial E(R)}}
   \Xzeroframe
    \left[
        \bcov{^i}
             \left(
              \backg_{J K} g^{J K}
             \right)
          -
         \bcov{_J } g^{J i}
    \right]
     d\spaceS_{i}
     \,.
      \label{23IV17.1+}
\end{equation}
Existence of the limit in \eqref{23IV17.1+} will be guaranteed if, instead of \eqref{30IV17.1}-\eq{30IV17.4}, one assumes now, e.g.,
\bel{30IV17.5}
 \int_{\hyp \cap \{r\ge R\}}
  |X |
   \Big(
    \sum_{\fr \mu \fr \alpha\fr \beta} |\bcov{_{\fr \mu}} g_{\fr \alpha\fr \beta}|^2
  + \sum_{ \fr \alpha\fr \beta} |T_{\fr \alpha\fr \beta}| + |\Lambda| |g^{\mu\nu}(g_{\mu\nu}-\backg_{\mu\nu})| \Big)
   d \mu_{\backg_\hyp} <\infty
  \,,
\ee
where $d \mu_{\backg_\hyp}$ is the $(n\pKKN )$-dimensional Riemannian measure induced on $\hyp$ by $\backg$.
A condition on the metric and the energy-momentum tensor of matter fields naturally associated with \eq{30IV17.5} is
\bel{30IV17.6}
  \lim_{R\to\infty} |X| |\partial E(R)| T_{\fr \mu\fr \nu}= 0
 \,,
 \qquad
  \lim_{R\to\infty} |X| |\partial E(R)| |\Lambda| |g^{\mu\nu}(g_{\mu\nu}-\backg_{\mu\nu})|  = 0
 \,,
\ee
where $ |\partial E(R)|$ denotes the area of $ \partial E(R) $; compare \eq{30IV17.4}. This will be assumed whenever relevant.

As an example, we consider the Rasheed metrics of Appendix~\ref{s14II17.6} with $P\ne 0$, which are vacuum. The $\backg$-Killing vector $X=\partial_t$ is $\backg$-covariantly constant so that \eqref{23IV17.1+} applies. The asymptotic behaviour of the metric coefficients in the frame \eq{19IV17.2} coincides with the asymptotic behaviour of the metric coefficients in manifestly asymptotically Minkowskian coordinates when $P=0$, and is given by \eq{31III17.11}. One obtains
\bel{31III17.12+}
 p_0 = 4\pi  P M
  \,,
\ee
where the extra factor $4P$, as compared to \eq{31III17.12}, is due to the $8P\pi$-periodicity of the coordinate $x^4$ (cf.~\eq{22IV17.7}), as enforced by the requirement of smoothness of the metric. Note that the formulae \eq{31III17.12+=a} for the ADM four-momentum remain unchanged.

We emphasise that the calculations above are done at fixed $P$, since every $P$ defines its own class of asymptotic backgrounds. As a result, the phase space of all configurations considered above splits into sectors parameterised by $P$.
It would be interesting to investigate the question of existence of a Hamiltonian in a phase space where $P$ is allowed to vary. We leave this question to future work.

If $X$ is not $\backg$-covariantly constant, the second term of \eqref{a.7} does not vanish. Thus, disregarding those terms which do \emph{not} involve the forms $\dScoordinate_{0i}$, we obtain, keeping in mind that $X$ is a Killing vector field of $\backg$,
\begin{eqnarray}
 \nn
  \lefteqn{ \frac 12
 ( U^{\alpha\beta} - U^{\alpha\beta}\big|_{g=\backg})  \dScoordinate_{\alpha\beta}
 =
  \frac 12 \big(  \tensor{\mathbb{U}}{^{\alpha\beta}_\lambda}X^\lambda
  -
  \frac{1}{8\pi}
   \sqrt{|\det g|} g^{\mu[\alpha}\delta^{\beta]}_\nu \overline \nabla_\mu  \tensor{X}{^\nu }
   - U^{\alpha\beta}\big|_{g=\backg}
    \big)
    \dScoordinate_{\alpha\beta}
}
&&
  \\
  \nn
  &=&
   \frac{3}{16\pi}
    \delta^{\alpha\beta\gamma}_{\lambda\mu\nu}
    X^\nu \backg^{\lambda\rho}\backg_{\gamma\sigma}\bcov{_\rho} g^{\sigma\mu}
    dS_{\alpha\beta}
  -
  \frac{1}{16\pi}
  \big(
   \sqrt{|\det g|} g^{\mu[\alpha}-
    \sqrt{|\det \backg|}
    \backg^{\mu[\alpha} \big) \bcov{_\mu} X^{\beta]} \dScoordinate_{\alpha\beta}
    +\nc
  \\
  \nn
  &=&
\frac{1   }{8\pi}
   \bigg(
 3
    \delta^{0 i \gamma}_{\lambda\mu\nu}
    X^\nu \backg^{\lambda\rho}\backg_{\gamma\sigma}\bcov{_\rho} g^{\sigma\mu}
  -
  \big(
   g^{\mu[0}- e^{-1}
     \backg^{\mu[  0} \big) \bcov{_\mu} X^{i]}
       \bigg) N
       d\spaceS_{i}
    +\nc
\end{eqnarray}
Here we have used $dS_{ 0i} = N d\spaceS_i$, where $N$ is the lapse function of the foliation by the level sets of $t$,  defined by writing the metric as $g=-N^2dt^2 + g_{IJ}(dx^I+N ^I dt)(dx^J+N ^Jdt)$. We conclude that
\begin{equation}
 \label{30IV17.32}
  \HboundX
   =
   \frac{1}{8\pi}
   \lim_{R\rightarrow\infty}\int_{\nored{\partial E(R)}}
   \Big(3  \delta^{0 i \gamma}_{\lambda\mu\nu}
   X^\nu\backg^{\lambda\rho}\backg_{\gamma\sigma}\bcov{_\rho} g^{\sigma\mu}
  -
   \big(
   { g^{\mu[0} }-
    e^{-1} \backg^{\mu  [0} \big) \bcov{_\mu} X^{i]}
    \Big) N
       d\spaceS_{i}\,.
\end{equation}
%
%

We can apply the last formula to the background Killing vectors $\partial_i$ and $\partial_4$ for Rasheed metrics with $P\ne 0$.
A  calculation gives
\bel{31III17.12+5}
  \quad
   p_i = 0
  \,,
  \quad
   p_4 = 4\pi  P  Q
 \,.
\ee
Here one can note that $\partial_z$ is $\backg$-covariantly constant so that the last term in \eq{30IV17.32} does certainly not contribute; while $p_x=p_y=0$ follows from the axi-symmetry of the Rasheed metrics. (In fact $\bcov X  = O(r^{-2})$ or better for these Killing vectors so that the last term never contributes in the current case.)

 \Eq{30IV17.32} applies for completely general background metrics $\backg$, assuming that \eq{30IV17.5} and \eq{19IV17.1+} hold,
  for a large class of field equations. In particular it applies to asymptotically Kottler (``anti-de Sitter'') metrics, compare~\cite{ChHerzlich,Wang,CJKKerrdS,ChruscielSimon}.

\section{Energy-momentum and the curvature tensor}
  \label{s14II17.3}

For our further purposes it is convenient to rewrite  \eq{14II17.2} in terms of the Christoffel symbols. As a first step towards this we note the following consequence of \eq{2.1}
\begin{equation}\label{2.2}
\delta^{\alpha\beta\gamma}_{\lambda\mu\nu}\,dS_{\alpha\beta}=\frac{1}{3}\cdot
\frac{1}{(n\pKKN -2)!}\epsilon_{\lambda\mu\nu {\xi_1}\cdots {\xi_{n\pKKN -2}}} \, dx^\gamma\wedge dx^{\xi_1}\wedge\cdots\wedge dx^{\xi_{n\pKKN -2}}\,.
\end{equation}
\subsection{KK-asymptotic  flatness}
 \label{ss21.IV.17}

We assume again that $X$ is $\backg$-covariantly constant; of course, it would suffice to assume that $\overline \nabla X$  falls-off fast enough to provide a vanishing contribution to the integral defining the Hamiltonian in the limit.

In the KK-asymptotically flat case \eq{a.12} can be rewritten as
\begin{equation}\label{2.3}
 p_\mu X^\mu_\infty=\frac{(-1)^{n\pKKN -1}}{16\pi (n\pKKN -2)!} \lim_{R\rightarrow \infty}
  \int_{S(R)\times \mathbb{T}^n}
   \epsilon_{\lambda\mu\nu {\xi_1}\cdots {\xi_{n\pKKN -2}}} X^\nu g^{\lambda\rho} \tensor{\Gamma }{^\mu_\gamma_\rho}\, dx^{\xi_1}\wedge\cdots\wedge  dx^{\xi_{n\pKKN -2}} \wedge dx^\gamma
    \,.
\end{equation}

In the standard asymptotically flat case, without Kaluza-Klein directions, \eq{2.3}  can be used to obtain an expression for the ADM energy-momentum in terms of the Riemann tensor, generalising a similar formula derived by Ashtekar and Hansen in space-time dimension four~\cite{AshtekarHansen} (compare~\cite{ChRemark,TanabeTanahashiShiromizu}), as follows: We can write
\begin{eqnarray}
\nonumber
 \lefteqn{
  \epsilon_{\lambda\mu\nu {\xi_1}\cdots {\xi_{n-2}}} X^\nu g^{\lambda\rho} \tensor{\nodiffGamma }{^\mu_\gamma_\rho}\,  dx^{\xi_1}\wedge\cdots\wedge  dx^{\xi_{n-2}}  \wedge dx^\gamma
  }
  &&
  \\[1em]
  \nonumber
  &=&
   d\left(\epsilon_{\lambda\mu\nu {\xi_1}\cdots {\xi_{n-2}}} X^\nu x^{\xi_1} g^{\lambda\rho} \tensor{\nodiffGamma }{^\mu_\gamma_\rho}\, dx^{\xi_2}\wedge\cdots\wedge  dx^{\xi_{n-2}} \wedge dx^\gamma\right)
   \\[1em]
   &&
    -
    (-1)^{n-3}\epsilon_{\lambda\mu\nu {\xi_1}\cdots {\xi_{n-2}}} X^\nu g^{\lambda\rho} x^{\xi_1} dx^{\xi_2}\wedge\cdots\wedge  dx^{\xi_{n-2}}\wedge \underbrace{\left(\partial_\sigma \tensor{\nodiffGamma }{^\mu_\gamma_\rho} dx^\sigma\wedge dx^\gamma\right)}_{=\frac{1}{2}\nodiffR{^\mu_\rho_\sigma_\gamma} dx^\sigma\wedge dx^\gamma}
    +
    \nc
  \label{7V17.31}
\end{eqnarray}
Inserting this into \eqref{2.3} and applying Stokes's theorem one obtains
%
\bean
 p_\mu X^\mu_\infty
  & = &
   \frac{1}{32\pi (n-2)!}\lim_{R\rightarrow\infty}\int_{S(R)} \epsilon_{\nu {\xi_1}\cdots {\xi_{n-2}}\lambda\mu} X^\nu x^{\xi_1} \nodiffR{^\lambda^\mu_\delta_\gamma}\,
    \underbrace{dx^{\xi_2}\wedge\cdots\wedge  dx^{\xi_{n-2}}\wedge dx^\delta\wedge dx^\gamma}_{- \frac 12 {\epsilon^{\xi_2\ldots \xi_{n-2}\delta\gamma\mu\nu}{}_ {dS_{\mu\nu}}}}
\\
 & = &
   \frac{1}{16  (n-2) \pi }\lim_{R\rightarrow\infty}\int_{S(R)} X^\mu x^\nu \, \nodiffR{_{\mu \nu \rho\sigma} } dS^{\rho\sigma}
   \,,
 \label{2.4a}
\eea
which is the desired new formula.

Let us now pass to a derivation of a version of \eqref{2.4a} relevant for Kaluza-Klein asymptotically flat space-times.
In this case we will be integrating
the integrand of \eqref{2.3} over
$$
 S^{n-1}\times  \mathbb{T}^\KKN
 =
 S^{d-2}\times \mathbb {T}^\KKN
 \,.
$$
So only those forms in the sum which contain a 
$dx^{d}\wedge \cdots \wedge dx^{d\pKKN-1 }$ factor will survive integration.
We will use the symbol

\begin{enumerate}
  \item
 $\nodiffKKR{^\alpha_{\beta\gamma\delta}}$ to denote the Riemann tensor of the $(d\pKKN )$-dimensional metric $g_{\mu\nu}dx^\mu dx^\nu $,

 \item $\nodiffsptR{^a_{bcd}}$ that of the $d$-dimensional metric $g_{ab}dx^a dx^b$,

 \item
 $\nodiffspKKR{^I_{JKL}}$ for the $(n\pKKN )$-dimensional metric $g_{IJ}dx^I dx^J$, and
 \item $\nodiffspR{^i_{jk\ell}}$ for that of the $n$-dimensional metric $g_{ij}dx^i dx^j$.
\end{enumerate}

No distinction between $g_{ab}dx^a dx^b$ and $g_{\mu\nu}dx^\mu dx^\nu $ will be made when $\KKN =0$.
Keeping in mind that $\nc$ denotes
the sum of those terms which do not contribute to the integral either because of the integration domain, or by Stokes theorem, or by passage to the limit, we find
\begin{eqnarray}
\nonumber 
  \lefteqn{
\epsilon_{\lambda\mu\nu {\xi_1}\ldots {\xi_{d\pKKN -3}}} X^\nu g^{\lambda\rho} \tensor{\nodiffGamma }{^\mu_\gamma_\rho}\,  dx^{\xi_1}\wedge\cdots\wedge  dx^{\xi_{d\pKKN -3}}  \wedge dx^\gamma
 }
 &&
\\
 \nn
 & = &
  \frac{(d\pKKN -3)!}{(d-3)! N! }
 \epsilon_{bc f {a_1}\ldots a_{d-3}A_1\ldots {A_{\KKN }}}
  X^f g^{b e} \tensor{\nodiffGamma }{^c_a_e}\,  dx^{a_1}\wedge\cdots\wedge  dx^{A_{\KKN }}  \wedge dx^a
\\
 \nn
 &&
 +
  \frac{(d\pKKN -3)!}{(d-2)!(N-1)!}
 \epsilon_{\lambda\mu\nu {a_1}\ldots a_{d-2}A_1\ldots {A_{\KKN -1}}}
  X^\nu g^{\lambda\rho} \tensor{\nodiffGamma }{^\mu_A_\rho}\,  dx^{a_1}\wedge\cdots\wedge  dx^{A_{\KKN -1}}  \wedge dx^A
\\
 \nn
\\
 \nn
 & = &
  \frac{(d\pKKN -3)!}{(d-3)!N!}
 d(\epsilon_{bc f {a_1}\ldots a_{d-3}A_1\ldots {A_{\KKN }}}
  X^f \eta^{b e} \tensor{\nodiffGamma }{^c_a_e}\,  x^{a_1} d x^{a_2}\wedge\cdots\wedge  dx^{A_{\KKN }}  \wedge dx^a)
\\
 \nn
 &&
 -
  \frac{(d\pKKN -3)!}{(d-3)!N!}
 \epsilon_{bc f {a_1}\ldots a_{d-3}A_1\ldots {A_{\KKN }}}
  X^f \eta^{b e} x^{a_1}d \tensor{\nodiffGamma }{^c_a_e}\wedge\cdots\wedge  dx^{A_{\KKN }}  \wedge dx^a
\\
 \nn
 &&
 +
   \frac{(d\pKKN -3)!}{(d-2)!(N-1)!}
 d(\epsilon_{\lambda\mu\nu {a_1}\ldots a_{d-2}A_1\ldots {A_{\KKN -1}}}
  X^\nu \eta^{\lambda\rho} \tensor{\nodiffGamma }{^\mu_A_\rho}\,  x^{a_1} d x^{a_2}\wedge\cdots\wedge  dx^{A_{\KKN -1}}  \wedge dx^A
  )
\\
 \nn
 &&
 -
   \frac{(d\pKKN -3)!}{(d-2)!(N-1)!}
 \epsilon_{\lambda\mu\nu {a_1}\ldots a_{d-2}A_1\ldots {A_{\KKN -1}}}
  X^\nu \eta^{\lambda\rho} x^{a_1} d \tensor{\nodiffGamma }{^\mu_A_\rho}\wedge\cdots\wedge  dx^{A_{\KKN -1}}  \wedge dx^A
 + \nc
 \\
 \nn
 &=&
    {
   \frac{(d\pKKN -3)!}{2(d-3)!N!}
 \epsilon_{bc f {a_1}\ldots a_{d-3}A_1\ldots {A_{\KKN }}}
  X^f   x^{a_1}  \nodiffKKR{^{b}^c_{a_{d-1}}_{a_{d-2}}} dx^{a_{d-1}} \wedge dx^{a_{2}} \wedge\cdots\wedge  dx^{A_{\KKN }} \wedge dx^{a_{d-2}}
   }
   \\
   &&
    {+}
   {
    \frac{(d\pKKN -3)!}{ (d-2)!(N-1)!}
 \epsilon_{\lambda\mu\nu {a_1}\ldots a_{d-2}A_1\ldots {A_{\KKN -1}}}
  X^\nu  x^{a_1} \nodiffKKR{^{\lambda}^\mu_{a_{d-1}}_{A}} dx^{a_{d-1}}\wedge dx^{a_{2}} \wedge
  \cdots\wedge  dx^{A_{\KKN -1}}  \wedge dx^A
 + \nc
   }
 \label{3VIII17.1}
\end{eqnarray}
%
Using
\beaa
\lefteqn{
  \epsilon_{\lambda\mu\nu {a_1}\ldots a_{d-2}A_1\ldots {A_{\KKN -1}}}  dx^{a_{d-1}}\wedge dx^{a_{2}} \wedge
  \cdots\wedge  dx^{A_{\KKN -1}}  \wedge dx^A
   }
\\
 &&
     =
  3 (N-1)! (-1)^{d\pKKN -1}\delta_{\lambda \mu \nu}^{A a b} \epsilon_{ab {a_1}\ldots a_{d-2} }  dx^{a_{d-1}}\wedge dx^{a_{2}}
   \wedge \cdots
   \wedge dx^{a_{d-2}} \wedge
  dx^{d+1}  \wedge\cdots   \wedge dx^{d\pKKN }
  \,,
\eeaa
after some reordering of indices one obtains
%
%
\begin{eqnarray}
   \nn
 p_\mu X^\mu_\infty
  & = &
    \frac{(-1)^n}{32\pi (n-1)!}
   \lim_{R\rightarrow\infty}\int_{S(R)} \int_{\mathbb{T}^\KKN }
   x^{a_1}
   \Big[
   (n-1) \epsilon_{a_1 a_2 \cdots a_{n-2}abc}\, X^a \nodiffKKR{^{bc}_{a_{n-1}}_{a_{n}}}
   \\
   &&
   -
   \underbrace{
     \epsilon_{a_1 a_2 \cdots a_{n-1}ab}
    \left(
    4 X^a \nodiffKKR{^{bA}_{a_{n}A}}+\redtwo X^A \nodiffKKR{^{ab}_{a_{n}A}}
     \right)
      }_{(*)}
   \Big]
    dx^{a_{2}}\wedge \cdots\wedge dx^{a_n}\wedge
    dx^{d+1}  \wedge\cdots   \wedge dx^{d\pKKN }
    \,.
     \label{26V17.11}
\end{eqnarray}
Using
\begin{eqnarray}
   dx^{a_{2}}\wedge \cdots\wedge dx^{a_n}\wedge dx^{d+1}  \wedge\cdots\wedge dx^{d\pKKN }
   =
   -\frac{1}{2}
   \epsilon^{a_2 \cdots a_n ef} dS_{ef}
\end{eqnarray}
and
\begin{eqnarray*}
  \nodiffKKR{^{ad}_{bd}}
  &=&
  \nodiffKKR{^{a}_{b}}-\nodiffKKR{^{aA}_{bA}}
\end{eqnarray*}
one obtains for the first term of the Hamiltonian integral, where in the fourth line below we use \eqref{2III17.4}, Appendix~\ref{sA2III17.1} below,
\begin{eqnarray}
  \nn
  \lefteqn{
   \epsilon_{a_1 a_2 \cdots a_{n-2}abc}\,
    x^{a_1} X^a  \nodiffKKR{^{bc}_{a_{n-1}}_{a_{n}}}
   dx^{a_{2}}\wedge \cdots\wedge dx^{a_n}\wedge
   dx^{d+1}  \wedge\cdots   \wedge dx^{d\pKKN }
   }
 \\
 \nn
 & = &
  -\frac{1}{2}
   \epsilon_{a_1 a_2 \cdots a_{n-2}abc}\,
   \epsilon^{a_2\cdots a_n ef}
    x^{a_1} X^a
   \nodiffKKR{^{bc}_{a_{n-1}}_{a_{n}}}
   dS_{ef}
   \\
   \nn
   &=&
   \frac{1}{2}
   (-1)^{n-1} (n-3)! \, 4!\,
   \delta^{a_{n-1}a_n ef}_{\,a_1 \, a\, b \, c}  x^{a_1} X^a
    \nodiffKKR{^{bc}_{a_{n-1}}_{a_{n}}}
    dS_{ef}
    \\
    \nn
    &=&
    2(-1)^{n-1} (n-3)! x^{a_1} X^a
    \left(
     \nodiffKKR{^{ef}_{a_1 a}}+\delta^{ef}_{a_1 a}
     \nodiffKKR{^{bc}_{bc}}-4\delta^{[e}_{[a_1} \nodiffKKR{^{f]c}_{a]c}}
    \right)
    dS_{ef}
    \\
    &=&
     2(-1)^{n-1} (n-3)! x^{a_1} X^a
     \left[
       \nodiffKKR{^{ef}_{a_1 a}}
       +
       \delta^{ef}_{a_1 a}
          \left(
           \nodiffKKR{^c_c}
           -
           \nodiffKKR{^{cA}_{cA}}
          \right)
      -
      4
          \left(
           \delta^{[e}_{[a_1} \nodiffKKR{^{f]}_{a]}}
            -
           \delta^{[e}_{[a_1} \nodiffKKR{^{f]A}_{a]A}}
          \right)
     \right]
     dS_{ef}
      \,.
\end{eqnarray}

Recall, now, that finiteness of the total energy of matter fields together with the dominant energy condition requires, essentially, that
\bel{10V17.11}
 T_{\mu\nu} = o(r^{-n})
 \,;
\ee
compare \eq{30IV17.4}.
This, together with the Einstein equations, implies
that the Ricci-tensor contribution to the integrals will vanish in the limit $R\to \infty$.   Nevertheless, we will keep the  Ricci tensor terms for future reference.

Using
\begin{eqnarray}
 \nn
  -\frac{1}{2}
  \epsilon_{a_1 a_2 \cdots a_{n-1}ab}\,
  \epsilon^{a_2 \cdots a_n ef} dS_{ef}
  =
  3 (-1)^n (n-2)!\,
  \delta^{a_n ef}_{a_1 ab} dS_{ef}
  \,,
\end{eqnarray}
the terms involving $(*)$ in \eq{26V17.11} can be manipulated as
\begin{eqnarray}
 \nn
 \lefteqn{
  \redsix  (-1)^n (n-2)!\,
  \delta^{a_n ef}_{a_1 ab} x^{a_1}
  \left(
   2X^a \nodiffKKR{^{bA}_{a_{n}A}}+X^A \nodiffKKR{^{ab}_{a_{n}A}}
  \right)
  dS_{ef}
 }
 \\
 \nn
 &=&
  \redtwo
  (-1)^n (n-2)!\, x^{a_1}
  \left(
   \delta^{a_n e}_{a_1 a} \delta^{f}_{b}
   +
   \delta^{e f}_{a_1a} \delta^{a_n}_{b}
   +
   \delta^{f a_n}_{a_1 a} \delta^{e}_{b}
  \right)
  \left(
   2X^a \nodiffKKR{^{bA}_{a_{n}A}}+X^A \nodiffKKR{^{ab}_{a_{n}A}}
  \right)
  dS_{ef}
  \\
  \nn
  &=& \redtwo
  (-1)^n (n-2)!
  \Big[
   2
    \left(
     x^{[a_n} X^{e]}\nodiffKKR{^{fA}_{a_n A}}
     +
     x^{[e} X^{f]}\nodiffKKR{^{a_n A}_{a_n A}}
     +
     x^{[f} X^{a_n]}\nodiffKKR{^{eA}_{a_n A}}
    \right)
    \\
    \nn
    &&
    +
    X^A
    \left(
    x^{[a_n} \nodiffKKR{^{e] f}_{a_n A}}
    +
    x^{[e} \nodiffKKR{^{f] a_n}_{a_n A}}
    +
    x^{[f} \nodiffKKR{^{a_n] e}_{a_n A}}
    \right)
  \Big]
  dS_{ef}
  \,.
\end{eqnarray}
%
Renaming the indices, rearranging terms, and plugging the results into the integral one obtains our final expression
\begin{eqnarray}
 \nn
   p_\mu X^\mu_\infty
  & = &
  \frac{1}{32\pi (n-2)}
  \times
  \\
  \nn
  &&
   \lim_{R\rightarrow\infty}\int_{S(R)} \int_{\mathbb{T}^\KKN }
   \Bigg\{
    -2x^{b} X^a
     \left[
       \nodiffKKR{^{ef}_{b a}}
       +
       \delta^{ef}_{b a}
          \left(
           \nodiffKKR{^c_c}
           -
           \nodiffKKR{^{cA}_{cA}}
          \right)
      -
      4
          \left(
           \delta^{[e}_{[b} \nodiffKKR{^{f]}_{a]}}
            -
           \delta^{[e}_{[b} \nodiffKKR{^{f]A}_{a]A}}
          \right)
     \right]
     \\
     \nn
     &&
     -
     \redtwo \frac{n-2}{n-1}
      \Bigg[
   2
    \left(
   2 x^{[b} X^{e]}\nodiffKKR{^{fA}_{b A}}
     +
      x^{e} X^{f}\nodiffKKR{^{bA}_{b A}}
    \right)
    +
    3 X^A
    x^{e} \nodiffKKR{^{fb}_{b A}}
  \Bigg]
  \Bigg\}
  dS_{ef}
  \\
  \nn
  &=&
   \frac{1}{16  (n-2) \pi}
   \lim_{R\rightarrow\infty}
   \int_{S(R)} \int_{\mathbb{T}^\KKN }
   \Bigg(
   X^a x^b \nodiffKKR{_{ab}^{ef}}
   +
   4 x^{[e}X^{a]} \nodiffKKR{^f_a}
     -
     x^e X^f \nodiffKKR{^c_c}
   \\
   &&
   -
    \frac{1}{n-1}
   \Big[
    (n-3) x^e X^f
     \nodiffKKR{^{bA}_{bA}}
    +
     4 x^{[e} X^{a]} \nodiffKKR{^{fA}_{aA}}
     +
     3(n-2)
     X^A
    x^{e} \nodiffKKR{^{fb}_{b A}}
    \Big]
    \Bigg)
    dS_{ef}
    \,. \label{31III17.5}
\end{eqnarray}

Some special cases are of interest:

\begin{enumerate}
  \item
  Suppose that $X^\mu=\delta^\mu_0$, thus $X$ has just time component. At $x^0=0$ we have
\begin{eqnarray}
 \nn
  p_0
  &=&
   \frac{1}{8 (n-2) \pi }
   \lim_{R\rightarrow\infty}
   \int_{S(R)} \int_{\mathbb{T}^\KKN }
   \Bigg[
    x^j \nodiffKKR{_{0j}^{0i}}
    +
     \frac 1 2
     x^i
         \left(
          \nodiffKKR{^j_j}-\nodiffKKR{^0_0}
         \right)
     -
     x^j \nodiffKKR{^i_j}
   \\
   &&
   -
    \frac{1}{n-1}
  \left(
       \frac 1 2
     x^{i} \nodiffKKR{^{0A}_{0A}}
    -
     x^{j} \nodiffKKR{^{iA}_{jA}}
     \right)
      \Bigg]
    dS_{0i}
    \,.
     \label{17IV17.2}
\end{eqnarray}
where the terms involving the Ricci tensor give a vanishing contribution in view of \eqref{10V17.11}; similarly for \eq{2III17.1} below.

  \item
   Suppose that $X^A=0$, thus $X$ has only space-time components. Then
\begin{eqnarray}
 \nn
   p_a X^a_\infty
  & = &
   \frac{1}{16  (n-2) \pi }
   \lim_{R\rightarrow\infty}
   \int_{S(R)} \int_{\mathbb{T}^\KKN }
   \Bigg(
   X^a x^b \nodiffKKR{_{a b}^{ef}}
   +
   4 x^{[e}X^{a]} \nodiffKKR{^f_a}
     -
     x^e X^f \nodiffKKR{^c_c}
   \\
   &&
   -
    \frac{1}{n-1}
   \Big[
     (n-3)x^e X^f
     \nodiffKKR{^{bA}_{bA}}
    +
    4x^{[e} X^{a]} \nodiffKKR{^{fA}_{aA}}
    \Big]
    \Bigg)
    dS_{ef}
    \,.
    \label{2III17.1}
\end{eqnarray}
We will see below that the first term in the right-hand side
 is related to the Komar integral. It is not clear whether or not the remaining terms vanish in general. However, when $X^0=0$, at $t=0$  the third term in the integrand
gives a vanishing contribution, so that the generators of space-translations read
\begin{eqnarray}
   p_i X^i_\infty
  & = &
   \frac{1}{8 (n-2) \pi }
   \lim_{R\rightarrow\infty}\int_{S(R)} \int_{\mathbb{T}^\KKN }
    \Bigg[
   X^i x^k \nodiffKKR{_{i k} ^{0j}}
    +
    2 x^{[i} X^{j]}
         \left(
         \frac{2}{  n-1 }
          \nodiffKKR{^{0A}_{iA}}
          +
          \nodiffKKR{^0_i}
         \right)
    \Bigg]
    dS_{0j}
   \,.
    \label{27III17.1}
\end{eqnarray}
We also note that when $\KKN =1$  the contribution of the fourth term in the integrand in \eq{2III17.1} always vanishes because then,
 denoting by $x^4$ the Kaluza-Klein coordinate,
$$
   \nodiffKKR{^{bA}_{bA}} =   \nodiffKKR{^{b 4}_{b 4}}=   \nodiffKKR{^{\mu 4}_{\mu 4}} =   \nodiffKKR{^{4}_{4}} = o(r^{-n})
   \,,\
$$
which gives a zero contribution in the limit.

 \item
   Suppose instead that $X^a=0$, thus $X$ has only  components tangential to the Kaluza-Klein fibers. Then, again at $x^0=0$,
\begin{eqnarray}
\nn
   p_A X^A_\infty
  & = &
    \frac{3}{16 (n-1) \pi}
   \lim_{R\rightarrow\infty}\int_{S(R)} \int_{\mathbb{T}^\KKN }
    X^A
    x^{e} \nodiffKKR{_{Ab}^{fb}}
   dS_{ef}
   \\
    &=&
    \frac{3}{16 (n-1)\pi}
   \lim_{R\rightarrow\infty}\int_{S(R)} \int_{\mathbb{T}^\KKN }
    X^A
    x^{i} \nodiffKKR{_{AB}^{0B}}
   dS_{0i}
   \,,
    \label{2III17.2}
\end{eqnarray}
where the decay $o(r^{-n})$
of the Ricci tensor of the $(n\pKKN +1)$-dimensional metric has been used.

\end{enumerate} 


\subsection{General case}
 \label{ss20IV17.2}

For general background metrics, still assuming a covariantly-constant $\backg$-Killing vector, we start by rewriting \eq{30IV17.31} as
\begin{equation}
 \label{2.3+}
  \HboundX
   =\frac{(-1)^{n\pKKN -1}}{16\pi (n\pKKN -2)!} \lim_{R\rightarrow \infty}
  \int_{\partial E(R)}
   \epsilon_{\lambda\mu\nu {\xi_1}\cdots {\xi_{n\pKKN -2}}} X^\nu g^{\lambda\rho} \tensor{\diffGamma }{^\mu_\gamma_\rho}\, dx^{\xi_1}\wedge\cdots\wedge  dx^{\xi_{n\pKKN -2}} \wedge dx^\gamma
    \,,
\end{equation}
where
\bel{3V17.11}
 \tensor{\diffGamma}{^{\alpha}_{\beta\gamma}} := \tensor{\Gamma}{^{\alpha}_{\beta\gamma}} - \tensor{\backGamma}{^{\alpha}_{\beta\gamma}} = o(r^{-\beta})
 \,,
\ee
with the last equality following from \eq{20VI17.1}.

In order to obtain a  version of  \eq{7V17.31} suitable to the current setting we  will assume that there exists a vector field $Z$ with $Z^A=0$
and  a real number  $\gamma>0$
such that
\bel{5V17.4}
 \bcov{_a} Z^b = \delta_a^b + O(r^{-\gamma}) \, \mod{}  ( \delta_\mu^r, \delta_\mu^t )
 \,.
\ee
Here we write ``$\!\!\!\!\mod{} ( \delta_\mu^r, \delta_\mu^t )$'' for a tensor which has the form
 $  \delta_{  \mu}^  r \alphaz_{\ldots}
   +  \delta_{  \mu}^  t \betaz_{\ldots}  $  for some tensor fields $\alphaz $, and $\betaz $ .
  That is to say, if $X$ is a vector field  tangent to the submanifolds of constant $t$, $r$,   and if ``$u_{\mu \ldots} = 0
 \mod{} ( \delta_\mu^r, \delta_\mu^t )$'', then $ X^\mu u_{\mu\ldots} = 0$.


We show in Appendix~\ref{A7V17.1} that the vector field defined in appropriate coordinates as
\bel{7V17.52}
 Z=r \partial_r
\ee
satisfies a) \eq{5V17.4} for asymptotically anti-de Sitter metrics, and b)  for general Rasheed metrics,  in both cases without the error term $O(r^{-\gamma})$; equivalently,  the exponent $\gamma$ can be taken as large as desired. We have introduced the $O(r^{-\gamma})$ term for possible future generalisations.

We further assume that
\bel{25VI17.1}
 \overline\nabla _\mu X^\nu =
 O(|X| r^{-\beta})
  \nored{
 \mod{} ( \delta_\mu^r, \delta_\mu^t )
 }
 \,,
\ee
which will certainly be the case if $X$ is $\backg$-covariantly constant.
Last but not least, we replace
 \eq{19IV17.1+} by the requirement that
\bean
 &&
   \mbox{terms $
  o\left(  |X| r^{-\alpha -\beta  }\right)$,  $
  o\left(|Z| |X| r^{-2\beta  }\right)$  and $
  o\left(|X| r^{- \beta -\gamma }\right)$ give vanishing contribution }
\\
 &&
    \mbox{to  boundary integrals at fixed $r$ and $t$, after passing to the limit $r\to \infty$.
     }
\eeal{19IV17.1++}

Now, the identity that we are about to derive will be integrated on submanifolds of fixed $r$ and $t$, so that any forms containing a factor $dr$ or $dt$ will give zero integral.  Assuming that there are no Kaluza-Klein directions ($\KKN =0$) we find
\begin{eqnarray}
\nonumber
 \lefteqn{  d\left(\epsilon_{\lambda\mu\nu {\xi_1}\cdots {\xi_{n-2}}} X^\nu Z^{\xi_1} g^{\lambda\rho} \tensor{\diffGamma }{^\mu_\gamma_\rho}\, dx^{\xi_2}\wedge\cdots\wedge  dx^{\xi_{n-2}} \wedge dx^\gamma\right)
 }
 &&
\\
\nn
 &=&  \overline \nabla{_\sigma} \left(\sqrt{|\det g|}\mathring \epsilon_{\lambda\mu\nu {\xi_1}\cdots {\xi_{n-2}}} X^\nu Z^{\xi_1} g^{\lambda\rho} \tensor{\diffGamma }{^\mu_\gamma_\rho}\right) dx^\sigma\wedge dx^{\xi_2}\wedge\cdots\wedge  dx^{\xi_{n-2}} \wedge dx^\gamma
\\
 \nn
 &=&  Z^{\xi_1}  \epsilon_{\lambda\mu\nu {\xi_1}\cdots {\xi_{n-2}}}
    g^{\lambda\rho} \tensor{\diffGamma }{^\mu_\gamma_\rho}
    \underbrace{\overline \nabla_{\sigma} X^\nu  dx^\sigma}_{\nc}
    \wedge
     dx^{\xi_2}\wedge\cdots\wedge  dx^{\xi_{n-2}} \wedge dx^\gamma
\\
  && +
  \epsilon_{\lambda\mu\nu {\xi_1}\cdots {\xi_{n-2}}} X^\nu g^{\lambda\rho} \tensor{\diffGamma }{^\mu_\gamma_\rho}
   \underbrace{
    \overline \nabla{_\sigma} Z^{\xi_1}\,  dx^{\sigma}
     }_{   dx^{\xi_1}  + \nc}
      \wedge
   dx^{\xi_2}   \wedge\cdots\wedge  dx^{\xi_{n-2}}  \wedge dx^\gamma
  \nonumber
\\
   \nonumber
   &&
    +
    (-1)^{n-3}\epsilon_{\lambda\mu\nu {\xi_1}\cdots {\xi_{n-2}}} X^\nu g^{\lambda\rho} \nored{Z^{\xi_1}} dx^{\xi_2}\wedge\cdots\wedge  dx^{\xi_{n-2}}\wedge
    \underbrace{
       \left( \overline \nabla{_\sigma} \tensor{\diffGamma }{^\mu_\gamma_\rho} dx^\sigma\wedge dx^\gamma\right)
        }_{=\big(\frac{1}{2}\diffR{^\mu_\rho_\sigma_\gamma} + o(r^{-2\beta})\big) \
     dx^\sigma\wedge dx^\gamma}
    + \nc
\\
 &=&
  \epsilon_{\lambda\mu\nu {\xi_1}\cdots {\xi_{n-2}}} X^\nu g^{\lambda\rho} \tensor{\diffGamma }{^\mu_\gamma_\rho}
    \,  dx^{\xi_1}\wedge\cdots\wedge  dx^{\xi_{n-2}}  \wedge dx^\gamma
  \nonumber
\\
   &&
    +
    (-1)^{n-3}\frac 12 \epsilon_{\lambda\mu\nu {\xi_1}\cdots {\xi_{n-2}}} X^\nu g^{\lambda\rho} \diffR{^\mu_\rho_\sigma_\gamma}   \, \nored{Z^{\xi_1}} dx^{\xi_2}\wedge\cdots\wedge  dx^{\xi_{n-2}}\wedge
     dx^\sigma\wedge dx^\gamma
   +\nc
  \label{10V17.1}
\end{eqnarray}

This identity replaces \eq{7V17.31} in the current setting. One can now repeat the remaining calculations of Section~\ref{ss21.IV.17}  by replacing every occurrence of the Christoffel symbols by the difference of those of $g$ and $\backg$, every occurrence of the Riemann tensor by the difference of the Riemann tensors of $g$ and $\backg$, and every occurrence of an undifferentiated $x^\alpha$ by $Z^\alpha$.
 Some care must be taken when generalising \eq{31III17.5} when passing from the background Riemann tensor to the background Ricci tensor because in \eq{31III17.5} all indices are lowered and raised with $g$. Thus, \eq{2III17.3} is replaced now by
%
%
\begin{eqnarray}
 3! \delta^{\sigma\gamma\alpha\beta}_{\lambda\mu\nu\xi}
 \left(
  \KKR{^\mu_{\rho\sigma\gamma}}-\bKKR{^\mu_{\rho\sigma\gamma}}
 \right)
 g^{\lambda\rho}
 &=&
  \left(
   \KKR{^{\alpha\beta}_{\xi\nu}}-\bKKR{^{[\alpha}_{\rho\xi\nu}} g^{\beta]\rho}
  \right)
  +
  \left(
   \KKR{}-\bKKR{_{\rho\lambda}}g^{\rho\lambda}
  \right) \delta^{\alpha\beta}_{\xi\nu}
   \nn
  \\
  &&
  -
   4 \delta^{[\alpha}_{[\xi} \KKR{^{\beta]}_{\nu]}}
  +
   2\delta^{[\alpha}_{[\xi} g^{\beta]\rho} \bKKR{_{\nu]\rho}}
    -
   2\bKKR{^{[\alpha}_{\rho\lambda[\xi}} \delta^{\beta]}_{\nu]} g^{\rho\lambda}
   \,.
\end{eqnarray}

The simplest situation is obtained when $\KKN =0$ so that $\NN$ is reduced to a point, and \eq{30IV17.32} becomes
\begin{eqnarray}
 \lefteqn{
  \HboundX
   =
   \frac{1}{8(n-2)\pi}
   \lim_{R\rightarrow\infty}\int_{\nored{ S(R)}}
   \Big\{ X^\nu Z^\xi
    \left(
     \noKKR{^{0i}_{\nu\xi}}
     -
     \bnoKKR{^{[0}_{\rho\nu\xi}} g^{i] \rho}
    \right)
    }
    &&
    \nn
    \\
    &&
    +
    X^{[0} Z^{i]}
    \left(
     \noKKR{}
     -
     \bnoKKR{_{\rho\lambda}} g^{\rho\lambda}
    \right)
   +
   2 X^{\nu} Z^{[0}  \noKKR{^{i]}_\nu}
  -
    2 Z^{\nu} X^{[0}  \noKKR{^{i]}_\nu}
    +
     ( Z^{\nu} X^{[0 } g^{i]\rho} - X^{\nu} Z^{[0 } g^{i]\rho})\bnoKKR{_{\nu\rho}}
     \nn
\\
 &&
    -
      X^{[\nu} Z^{i]} \bnoKKR{^0_{\rho\lambda\nu}} g^{\rho\lambda}
      +
      X^{[\nu} Z^{0]} \bnoKKR{^i_{\rho\lambda\nu}} g^{\rho\lambda}
    -  (n-2)
    \big(
   { g^{\mu[0} -} e^{-1}
    \backg^{\mu[   0} \big) \bcov{_\mu} X^{i]}
    \Big)
    \Big \}
    N d\sigma_i
     \,.
      \label{19V17.12}
\end{eqnarray}

\subsubsection{$\Lambda \ne 0$}
 \label{ss23VI17.2}

We wish to analyse \eq{19V17.12} for metrics $g$ which asymptote a maximally symmetric background $\backg$ with  $\Lambda \ne 0$. This case requires separate attention as then the background curvature tensor does not approach zero as we recede to infinity.
We note that the calculations in this section are formally correct independently of the sign of $\Lambda$, but to the best of our knowledge they are only relevant in the case $\Lambda <0$.

It is useful to decompose  the Riemann tensor into its irreducible components,
\beaa
 R_{\alpha  \beta  \gamma  \delta  }
  & = &
   W_{\alpha  \beta  \gamma \delta  } +\frac{1}{d-2}(R_{\alpha  \gamma }g_{ \beta  \delta  }-R_{\alpha  \delta  }g_{ \beta  \gamma }+R_{ \beta  \delta  }g_{\alpha  \gamma }-R_{ \beta \gamma }g_{\alpha  \delta  })
 -\frac{R}{(d-1)(d-2)}(g_{ \beta  \delta  }g_{\alpha  \gamma }-g_{ \beta  \gamma }g_{\alpha  \delta  })
\\
  & = &
   W_{\alpha  \beta  \gamma \delta  }
    +
     \frac{1}{d-2}(P_{\alpha  \gamma }g_{ \beta  \delta  }-P_{\alpha  \delta  }g_{ \beta  \gamma }+P_{ \beta  \delta  }g_{\alpha  \gamma }-P_{ \beta \gamma }g_{\alpha  \delta  })
  +
   \frac{R}{d (d-1) }(g_{ \beta  \delta  }g_{\alpha  \gamma }-g_{ \beta  \gamma }g_{\alpha  \delta  })
 \,,
\eeaa
where $W_{\alpha  \beta  \gamma  \delta  }$ is the Weyl tensor and $P_{\alpha \beta }$ the trace-free part of the Ricci tensor,
$$
 P_{\mu\nu} = R_{\mu\nu} - \frac R d g_{\mu\nu}
 \,.
$$
This leads to the following rewriting of  \eq{19V17.12}
\begin{eqnarray}
  \HboundX
  &   = &
   \frac{1}{8(n-2)\pi}
   \lim_{R\rightarrow\infty}\int_{\boundaryset }
   \Big\{ X^\nu Z^\xi
    \left(
     \noKKW{^{0i}_{\nu\xi}}
     -
      \bnoKKW{^{[0}_{\rho\nu\xi}}g^{i]\rho}
     +
      \frac{2 \noKKR{}}{n(n+1)}
     \delta^{[0}_{[\nu} \delta^{i]}_{\xi]}
     -
     \frac{2\bnoKKR{}}{n(n+1)} \delta ^{[0}_{[\nu} \backg_{\xi]\rho} g^{i] \rho}
    \right)
    \nn
\\
 &&
  +
    X^{[0} Z^{i]}
    \left(
     \noKKR{}
     -
     \bnoKKR{_{\rho\lambda}} g^{\rho\lambda}
    \right)
   +
     2 X^{\nu} Z^{[0}  \noKKR{^{i]}_\nu}
  -
    2 Z^{\nu} X^{[0}  \noKKR{^{i]}_\nu}
    +
     ( Z^{\nu} X^{[0 } g^{i]\rho} - X^{\nu} Z^{[0 } g^{i]\rho})\bnoKKR{_{\nu\rho}}
     \nn
\\
    &&
    -
    \frac{2\bnoKKR{}}{n(n+1)}
     \big(
      X^{[\nu} Z^{i]} \delta^0_{[\lambda}\backg_{\nu]\rho}
      -
      X^{[\nu} Z^{0]}  \delta^i_{[\lambda}\backg_{\nu]\rho}
       \big) g^{\rho\lambda}
    - (n-2)
    \big(
   { g^{\mu[0} - e^{-1}}
    \backg^{\mu[   0} \big) \bcov{_\mu} X^{i]}
    \Big\}
    N d\sigma_i
     \,.
      \label{19V17.11}
\end{eqnarray}
Assuming that  the background Weyl tensor falls-off sufficiently fast so that it does not contribute to the integrals (e.g., vanishes, when the background is a space-form such as the anti-de Sitter metric),
and that both the energy-momentum tensor of matter and $e-1$ decay fast enough (cf.~\eq{30IV17.6}),
and setting
$$
 \Delta^{\mu\nu}:= g^{\mu\nu}-\backg^{\mu\nu}
 \,,
$$
we obtain
\begin{eqnarray}
 \lefteqn{
  \HboundX
   =
   \frac{1}{8(n-2)\pi}
   \lim_{R\rightarrow\infty}\int_{\boundaryset }
   \Big\{ X^\nu Z^\xi
    \left(
     \noKKW{^{0i}_{\nu\xi}}
     -
     \frac{2\bnoKKR{}}{n(n+1)} \delta ^{[0}_{[\nu} \backg_{\xi]\rho} \Delta^{i] \rho}
    \right)
    }
    &&
    \nn
\\
 &&
 -
    X^{[0} Z^{i]}
     \bnoKKR{_{\rho\lambda}} \Delta^{\rho\lambda}
    +
     ( Z^{\nu} X^{[0 } \Delta^{i]\rho} - X^{\nu} Z^{[0 } \Delta^{i]\rho})\bnoKKR{_{\nu\rho}}
     \nn
\\
    &&
    -
    \frac{2\bnoKKR{}}{n(n+1)}
     \big(
      X^{[\nu} Z^{i]} \delta^0_{[\lambda}\backg_{\nu]\rho}
      -
      X^{[\nu} Z^{0]}  \delta^i_{[\lambda}\backg_{\nu]\rho}
       \big) \Delta^{\rho\lambda}
    - (n-2)
    \Delta^{\mu[  0}  \bcov{_\mu} X^{i]}
    \Big\}
    N d\sigma_i
     \,.
      \label{22V17.1}
\end{eqnarray}
where we have also used the hypothesis \eq{19IV17.1+}
that terms such as $|X||Z| \Delta^{\mu\nu}\Delta_{\rho\sigma}$ and $|X||Z| g_{\mu\nu}\Delta^{\mu\nu}$ fall off fast enough so that they give no contribution to the integral in the limit.
With some further work one gets
\begin{eqnarray}
  \HboundX
  & =&
   \frac{1}{8(n-2)\pi}
   \lim_{R\rightarrow\infty}\int_{\boundaryset }
   \Big\{ X^\nu Z^\xi
     \noKKW{^{0i}_{\nu\xi}}
  \nn
\\
 &&
     +
       (n-2)\Delta ^{\mu[0}
      \left[
    \frac{  \bnoKKR{}}{n(n+1)}
      \left( X_\mu Z^{i]} - Z_\mu X^{i]}\right)
    -
      \bcov{_\mu} X^{i]}
    \right]
    \Big\}
    N d\sigma_i
     \,.
      \label{24V17.11}
\end{eqnarray}

 To continue, we assume the Birmingham-Kottler form \eq{24V17.21}-\eq{24V17.22} of the background metric $\backg$.
If $X$ is the $\backg$-Killing vector field $\partial_t$ then, writing momentarily $X_\nu$ for $\backg_{\nu\mu}X^\nu$
$$
 \overline \nabla_\sigma X_\nu  dx^\sigma \otimes dx^\nu =
 \overline \nabla_{[\sigma} X_{\nu]}  dx^\sigma \otimes dx^\nu =
  \partial_{[\sigma} X_{\nu]}  dx^\sigma \otimes dx^\nu =
  \partial_{[\sigma} \backg_{\nu]0}  dx^\sigma \otimes dx^\nu =
   \frac 12
  \partial_{ r} \backg_{0 0} dx^r \wedge dx^0 =
   \frac 12
  \partial_{ r} \backg_{0 0} \overline{\Theta}^1 \wedge \overline{\Theta}^0
  \,.
$$
Using this one checks that all terms linear in $\Delta$ in  \eq{24V17.11} cancel out, leading to the elegant formulae
\begin{eqnarray}
  \HboundX
    &= &
   \frac{1}{8(n-2)\pi}
   \lim_{R\rightarrow\infty}\int_{\boundaryset }  X^\nu Z^\xi
     \noKKW{^{0i}_{\nu\xi}}
    N d\sigma_i
     \nn
\\
    &= &
   \frac{1}{16(n-2)\pi}
   \lim_{R\rightarrow\infty}\int_{\boundaryset }  X^\nu Z^\xi
     \noKKW{^{\alpha\beta}_{\nu\xi}}
     dS_{\alpha\beta}
     \,,
           \label{24V17.222}
\end{eqnarray}
which, at this stage, hold  for all $X$ belonging to the $(n+1)$-dimensional family of Killing vectors of the anti-de Sitter background which are normal to $\{t=0\}$.

If $X=\partial_\varphi$, then we have
\begin{eqnarray*}
 \bcov{_\sigma} X_\nu  dx^\sigma \otimes dx^\nu
 &=&
 \bcov{_{[\sigma}} X_{\nu]}  dx^\sigma \otimes dx^\nu =
  \frac{1}{2} \partial_\sigma \backg{_{\nu\varphi}} dx^\sigma \wedge dx^\nu =
   \frac{1}{2} \partial_r \backg_{\varphi\varphi} dr\wedge d\varphi
   +
   \frac{1}{2} \partial_\theta \backg{_{\varphi\varphi}} d\theta\wedge d \varphi
\\
   &=&
   \sqrt{V} \sin\theta \, \overline{\Theta}^1 \wedge \overline{\Theta}^3
   +
   \cos\theta \, \overline{\Theta}^2 \wedge \overline{\Theta}^3
   \,,
\end{eqnarray*}
where we used the co-frame of the background metric (B.1) with the following co-basis
\begin{equation}
 \overline{\Theta}^0 = \sqrt{V} dt\,,
 \quad \overline{\Theta}^1 = \frac{1}{\sqrt{V}} dr\,, \quad
  \overline{\Theta}^2 = r d\theta \,, \quad
  \overline{\Theta}^3 = r \sin\theta d\varphi\,.
\end{equation}
Hence, in this co-frame one obtains
\begin{eqnarray*}
 &
 \bcov{_{\fr1}} X_{\fr3} = \sqrt{V} \sin\theta = -\bcov{_{\fr3}} X_{\fr1}\,,
  \qquad
 \bcov{_{\fr2}} X_{\fr3} =
  \cos\theta = -\bcov{_{\fr3}} X_{\fr2}\,.
   &
\end{eqnarray*}
Therefore, the second term of the integrand in \eq{24V17.11} vanishes for $r\longrightarrow\infty$, since (keeping in mind that $dS_{\fr0\fr i}$ for $i\neq 1$  gives zero contribution to the integrals)

 \begin{eqnarray*}
 \lefteqn{
 \frac{\bnoKKR{}}{n(n+1)}
  \Delta^{\mu [0}
   \left(
    X_\mu Z^{1]} - Z_\mu X^{1]}
   \right)
   -
   \Delta^{\mu[0} \bcov{_\mu} X^{1]}
   =
   -\frac{\lambda}{2} \Delta^{\fr \mu \fr0} X_{\fr \mu} Z^{\fr1}
   -
   \frac{1}{2} \Delta^{\fr \mu \fr 0} \bcov{_\fr\mu} X^{\fr1}
   }
   &&
   \\
   &&
   =
   -\frac{\lambda}{2} \Delta^{\fr 3 \fr0} \underbrace{X_{\fr3}}_{=\left(\backg_{\fr3\fr3}+\Delta_{\fr3\fr3}\right)X^{\fr3}} Z^{\fr1}
   -
    \frac{1}{2} \Delta^{\fr 3 \fr 0} \bcov{_\fr3} X_{\fr1}
    =
    \left(
   - \frac{\lambda}{2} \cdot \frac{r^2 }{\sqrt{V}}
    +
    \frac{1}{2} \sqrt{V}
    \right) \sin\theta \Delta^{\fr 3 \fr0}
    \xrightarrow{r\rightarrow\infty} 0
    \,.
 \end{eqnarray*}	
Hence \eq{24V17.222} also holds for $X=\partial_\varphi$. Since all Killing vectors of AdS space-time can be obtained as linear combinations of images   of these two vectors by isometries preserving $\{t=0\}$, we conclude that  \eq{24V17.222} holds for all Killing vectors of the AdS metric.

Once this work was completed we have been informed that \eq{24V17.222} has already been observed in~\cite{AvdBP}, following-up on the pioneering definitions in~\cite{AshtekarMagnonAdS,AshtekarDas}. We note that our conditions for the equality in \eq{24V17.222} are quite weaker than those in~\cite{AvdBP}.

\subsubsection{$\Lambda = 0$}
 \label{ss23VI17.1}

We pass to the case $\Lambda =0$.
We will impose conditions which guarantee that   all terms which are quadratic or higher in $g_{\mu\nu}-\backg_{\mu\nu}$ give zero contribution to the integrals in the limit $R\to \infty$.
 Without these assumptions the final formulae become unreasonably long. Hence we assume \eq{3V17.11}, \eq{5V17.4}, \eq{25VI17.1} and \eq{19IV17.1++}.

In the current context, the calculation \eqref{3VIII17.1} is replaced by
\begin{eqnarray*}
  \lefteqn{
\epsilon_{\lambda\mu\nu {\xi_1}\ldots {\xi_{d\pKKN -3}}} X^\nu g^{\lambda\rho} \tensor{\diffGamma }{^\mu_\gamma_\rho}\,  dx^{\xi_1}\wedge\cdots\wedge  dx^{\xi_{d\pKKN -3}}  \wedge dx^\gamma
 }
 &&
\\
 & = &
 \frac{(d\pKKN -3)!}{(d-3)! N! }
 \epsilon_{bc f {a_1}\ldots a_{d-3}A_1\ldots {A_{\KKN }}}
  X^f g^{b e} \tensor{\diffGamma }{^c_a_e}\,  \underbrace{dx^{a_1}}_{\bcov{_h} Z^{a_1} dx^h +\nc=\delta^{a_1}_{h} dx^h+\nc}\wedge\cdots\wedge  dx^{A_{\KKN }}  \wedge dx^a
\\
 &&
 +
  \frac{(d\pKKN -3)!}{(d-2)!(N-1)!}
 \epsilon_{\lambda\mu\nu {a_1}\ldots a_{d-2}A_1\ldots {A_{\KKN -1}}}
  X^\nu g^{\lambda\rho} \tensor{\diffGamma }{^\mu_A_\rho}\,  dx^{a_1}\wedge\cdots\wedge  dx^{A_{\KKN -1}}  \wedge dx^A
\\
& = &
 \frac{(d\pKKN -3)!}{(d-3)!N!}
  \Bigg[
 \bcov{_h} (\epsilon_{bc f {a_1}\ldots a_{d-3}A_1\ldots {A_{\KKN }}}
  X^f g^{b e} \tensor{\diffGamma }{^c_a_e}\,  Z^{a_1}) dx^h \wedge d x^{a_2}\wedge\cdots\wedge  dx^{A_{\KKN }}  \wedge dx^a
  \\
  &&
  -
  \epsilon_{bc f {a_1}\ldots a_{d-3}A_1\ldots {A_{\KKN }}}
   \underbrace{\bcov{_h}  X^f}_{\nc}  g^{b e} \tensor{\diffGamma }{^c_a_e}\,  Z^{a_1} dx^h \wedge d x^{a_2}\wedge\cdots\wedge  dx^{A_{\KKN }}  \wedge dx^a
   \\
   &&
   -
   \epsilon_{bc f {a_1}\ldots a_{d-3}A_1\ldots {A_{\KKN }}}
   X^f g^{b e} Z^{a_1} \bcov{_h} \tensor{\diffGamma }{^c_a_e}\,   dx^h \wedge d x^{a_2}\wedge\cdots\wedge  dx^{A_{\KKN }}  \wedge dx^a
   +
   \nc
   \Bigg]
   \\
   &&
  +
   \frac{(d\pKKN -3)!}{(d-2)!(N-1)!}
   \Bigg[
 \bcov{_h} (\epsilon_{\lambda\mu\nu {a_1}\ldots a_{d-2}A_1\ldots {A_{\KKN -1}}}
  X^\nu g^{\lambda\rho} \tensor{\diffGamma }{^\mu_A_\rho}\,  Z^{a_1}) dx^h \wedge d x^{a_2}\wedge\cdots\wedge  dx^{A_{\KKN -1}}  \wedge dx^A
  \\
  &&
  -
  \epsilon_{\lambda\mu\nu {a_1}\ldots a_{d-2}A_1\ldots {A_{\KKN -1}}}
  \underbrace{\bcov{_h} X^\nu}_{\nc} g^{\lambda\rho} \tensor{\diffGamma }{^\mu_A_\rho}\,  Z^{a_1} dx^h \wedge d x^{a_2}\wedge\cdots\wedge  dx^{A_{\KKN -1}}  \wedge dx^A
  \\
  &&
  -
  \epsilon_{\lambda\mu\nu {a_1}\ldots a_{d-2}A_1\ldots {A_{\KKN -1}}}
  X^\nu g^{\lambda\rho} Z^{a_1} \bcov{_h} \tensor{\diffGamma }{^\mu_A_\rho}  dx^h \wedge d x^{a_2}\wedge\cdots\wedge  dx^{A_{\KKN -1}}  \wedge dx^A
  +
  \nc
  \Bigg]
  \\
  &=&
  \frac{(d\pKKN -3)!}{(d-3)!N!}
 d (\epsilon_{bc f {a_1}\ldots a_{d-3}A_1\ldots {A_{\KKN }}}
  X^f g^{b e} \tensor{\diffGamma }{^c_a_e}\,  Z^{a_1} d x^{a_2}\wedge\cdots\wedge  dx^{A_{\KKN }}  \wedge dx^a)
  \\
  &&
  +
   \frac{(d\pKKN -3)!}{(d-2)!(N-1)!}
 d (\epsilon_{\lambda\mu\nu {a_1}\ldots a_{d-2}A_1\ldots {A_{\KKN -1}}}
  X^\nu g^{\lambda\rho} \tensor{\diffGamma }{^\mu_A_\rho}\,  Z^{a_1} dx^h \wedge d x^{a_2}\wedge\cdots\wedge  dx^{A_{\KKN -1}}  \wedge dx^A)
  \\
  &&
  -
    \frac{(d\pKKN -3)!}{(d-3)!N!}
    \epsilon_{bc f {a_1}\ldots a_{d-3}A_1\ldots {A_{\KKN }}}
  X^f   x^{a_1} g^{be}  \diffKKR{^c_e_{a_{d-1}}_{a_{d-2}}} dx^{a_{d-1}} \wedge dx^{a_{2}} \wedge\cdots\wedge  dx^{A_{\KKN }} \wedge dx^{a_{d-2}}
    \\
    &&
    -
    \frac{(d\pKKN -3)!}{2(d-3)!N!}
    \epsilon_{\lambda\mu\nu {a_1}\ldots a_{d-2}A_1\ldots {A_{\KKN -1}}}
  X^\nu  x^{a_1} g^{\lambda\rho} \diffKKR{^\mu_\rho_{a_{d-1}}_{A}} dx^{a_{d-1}}\wedge dx^{a_{2}} \wedge
  \cdots\wedge  dx^{A_{\KKN -1}}  \wedge dx^A
  +
  \nc
    \,.
\end{eqnarray*}
 As before, in the last  equality we have used the fact that the first $\bcov{_h}$-terms in the first expression in each of the square brackets can be replaced by $\bcov {_\mu}$, because each form appearing in the first line above must already contain $\KKN$-factors of the $KK$ differentials $dx^A$, otherwise it will give zero contribution to the integral.

In addition to all the hypotheses so far we will also assume that the Riemann tensor decays at a rate $o(r^{-\beta_R})$:
\bel{24V17.31}
  \KKR{^\alpha_{\beta\gamma\delta}} = o (r^{-\betaR})
  \,,
  \quad
    \bKKR {^\alpha_{\beta\gamma\delta}} = o (r^{-\betaR})
  \,,
\ee
with $\betaR$ chosen so that
\bel{25VI17.11}
 \mbox{terms $|X||Z| o(r^{-\alpha-\betaR})$ give no contribution to the integral in the limit $R\to\infty$.}
\ee

All these conditions are satisfied by the five-dimensional Rasheed metrics, with $\alpha>0$ as close to one as one wishes, $\beta = 1+\alpha$, $\betaR =3$, with $\gamma$ 
as large as desired.

In line with our previous notation, we will write $\dKKR{^\alpha_{\beta\gamma\delta}}$ for the difference of Riemann tensors of the $(d\pKKN )$-dimensional metrics $g_{\mu\nu}dx^\mu dx^\nu $ and $\backg_{\mu\nu} dx^\mu dx^\nu$, $\dsptR{^a_{bcd}}$ for that of the $d$-dimensional metrics $g_{ab}dx^a dx^b$ and $\backg_{ab}dx^a dx^b$,
 $\dspKKR{^I_{JKL}}$ for the $(n\pKKN )$-dimensional metrics $g_{IJ}dx^I dx^J$ and $\backg_{IJ}dx^I dx^J$, and $\dspR{^i_{jk\ell}}$ for that of the $n$-dimensional metrics $g_{ij}dx^i dx^j$ and $\backg_{ij}dx^i dx^j$.
%

With the above hypotheses, the derivation of the key formula \eq{31III17.5} follows closely the remaining calculations in Section~\ref{ss21.IV.17},
 and leads to
\begin{eqnarray}
 \nn
  \lefteqn{
  \nored{ \HboundX }
  =
   \frac{1}{16  (n-2) \pi }
   \lim_{R\rightarrow\infty} \Bigg\{
    \int_{\partial E(R)}
   \Bigg(
   X^a \nored{Z}^b (\dKKR{_{ab}^{ef}})
   +
    4 \nored{Z}^{[e}X^{a]} \left( \dKKR{^f_a} \right)
   }
   &&
   \\
   \nn
   &&
   -
     \nored{Z}^e X^f \left( \dKKR{^c_c} \right)
   -
    \frac{1}{n-1}
   \Big[
    (n-3) \nored{Z}^e X^f
     (\dKKR{^{bA}_{bA}})
    +
     4 \nored{Z}^{[e} X^{a]} (\dKKR{^{fA}_{aA}})
    \\
    &&
    +
    3 (n-2)
    X^A
    \nored{Z}^{e}
     (\dKKR{^{fb}_{b A}})
  \Big]
  \Bigg)
   dS_{ef}
   -
   (n-2)
   \int_{\nored{\partial E(R)}}
    \big(
   { g^{\mu[a} - e^{-1}}
    \backg^{\mu[a} \big) \bcov{_\mu} X^{b  ]}
      dS_{ab}
    \Bigg\}
    \,.
     \label{31III17.5+}
\end{eqnarray}
For Rasheed solutions, or more generally for solutions which asymptote to the Rasheed backgrounds $\backg$ given by \eq{11IV17.1} with the usual decay $o(r^{-(n-2)/2})$, with $T_{\mu\nu} = o(r^{-3})$,  one has (cf.\ \eq{19V17.2}-\eq{19V17.1}) $\bKKR{^0_{\lambda\mu\nu}}=0$, $\bKKR{^ {  \fr \mu}}{} _{\fr  4 \fr \alpha \fr 4}=O(r^{-4})$,
 $\bKKR{_ {  \mu   \nu}}=O(r^{-4})$  and we obtain, for $X = \partial_\mu$
  and after passing to the limit $R\to \infty$, an integrand which is formally identical to that for metrics which are KK-asymptotically flat:
\be
  \HboundX
  =
   \frac{1}{16(n-2)\pi}
   \lim_{R\rightarrow\infty}\int_{\nored{  S(R)\times S^1}}
   X^\nu Z^\mu
     \KKR{^{\alpha\beta}_{\nu \mu}}
    dS_{\alpha\beta}
     =
   \frac{1}{8(n-2)\pi}
   \lim_{R\rightarrow\infty}\int_{\nored{  S(R)\times S^1}}
   X^\nu x^j
     \KKR{^{0i}_{\nu j}}
    N d\sigma_i
    \,.
\ee

Some special cases, without necessarily assuming that $g$ asymptotes to the Rasheed background, are of interest:

\begin{enumerate}
 \item
  Suppose that $X^\mu=\delta^\mu_0$, thus $X$ has just a time component. Keeping in mind that $Z^0=0$ and  $\partial E(R) \subset \{x^0=0\}$ we have
\begin{eqnarray}
 \nn
  \nored{ \Hb \left(\partial_0,\mcS \right)}
  &=&
   \frac{1}{8 (n-2) \pi  }
   \lim_{R\rightarrow\infty}
    \Bigg\{
     \int_{\partial E(R)}
     \Bigg(
    \nored{Z}^j \left(\dKKR{_{0j}^{0i}}\right)
    +
    \frac 1 2
     \nored{Z}^i
         \left(
           \KKR{^j_j}-\KKR{^0_0}
          +
         \nored{ \bKKR{^j_j}}-\nored{\bKKR{^0_0}}
         \right)
   \\
   \nn
   &&
     -
      \nored{Z}^j \left( \dKKR{^i_j} \right)
   -
    \frac{1}{2(n-1)}
       \Big[
     \nored{Z}^{i} (\dKKR{^{0A}_{0A}})
    -
    2
     \nored{Z}^{j}
     (\dKKR{^{iA}_{jA}} )
    \Big]
    \Bigg)
    dS_{0i}
\\
 \nn
  &&       -
   (n-2)
   \int_{\nored{\partial E(R)}}
   \big(
   { g^{\mu[0} - e^{-1}}
    \backg^{\mu[0} \big) \bcov{_\mu} X^{i  ]}
     dS_{0i}
     \Bigg\}
    \\
    \nn
    &=&
    \frac{1}{8 (n-2) \pi }
   \lim_{R\rightarrow\infty}
    \Bigg\{
     \int_{\partial E(R)}
      \Bigg(
       \frac{1}{2}
        \nored{Z}^i \left( \dKKR{^j_j} \right)
       -
        \nored{Z}^j  \left( \dKKR{_{Ij}^{Ii}} \right)
      \\
 \nn
      &&
      -
    \frac{1}{2(n-1)}
       \Big[
     \nored{Z}^{i} (\dKKR{^{0A}_{0A}})
    -
    2
     \nored{Z}^{j}
     (\dKKR{^{iA}_{jA}} )
    \Big]
    \Bigg)
    dS_{0i}
    \nn
\\
  &&
    -
  (n-2)
   \int_{\nored{\partial E(R)}}
   \big(
   { g^{\mu[0} - e^{-1}}
    \backg^{\mu[0} \big) \bcov{_\mu} X^{i  ]}
     dS_{0i}
     \Bigg\}
     \,.
     \label{17IV17.2+}
\end{eqnarray}
 \item
   Suppose that $X^A=0$, thus $X$ has only space-time components. Then
\begin{eqnarray}
 \nn
  \lefteqn{
  \nored{ \HboundX }
 =
   \frac{1}{16  (n-2) \pi }
   \lim_{R\rightarrow\infty}
   \Bigg\{ \int_{\partial E(R)}
   \Bigg(
   X^a \nored{Z}^b (\dKKR{_{a b}^{ef}})
   +
       4 \nored{Z}^{[e}X^{a]}
           \left(\dKKR{^f_a}\right)
   }
   &&
   \\
   \nn
   &&
       -
       \nored{Z}^e X^f \left( \dKKR{^c_c} \right)
   -
    \frac{1}{ n-1}
   \Big[
     (n-3) \nored{Z}^e X^f
     (\dKKR{^{bA}_{bA}})
    +
    4 \nored{Z}^{[e} X^{a]}
     (\dKKR{^{fA}_{aA}})
    \Big]
    \Bigg)
    dS_{ef}
    \nn
\\
  &&       -
   \nored{
  2 (n-2)
  \int_{\nored{\partial E(R)}}
   \big(
   { g^{\mu[0} - e^{-1}}
    \backg^{\mu[0} \big) \bcov{_\mu} X^{i  ]}
     dS_{0i}
    }
     \Bigg\}
   \,.
    \label{2III17.1+}
\end{eqnarray}
We will see below that the first term in the right-hand side
 is related to the Komar integral. It is not clear whether or not the remaining terms vanish in general. However, when $X^0=0$, at $t=0$  the third and fourth terms in the integrand
 in \eqref{2III17.1+} give a vanishing contribution so that the generators of space-translations read%
\begin{eqnarray}
 \nn
  \nored{ \HboundX }
  & = &
   \frac{1}{8 (n-2) \pi  }
   \lim_{R\rightarrow\infty}  \Bigg\{
   \int_{\partial E(R)}
   \Bigg(
   X^i \nored{Z}^k
    \left( \dKKR{_{i k} ^{0j}}\right)
    \\
    \nn
    &&
    +
    \nored{Z}^{[i} X^{j]}
    \left[
     \frac{2}{n-1}
     \left( \dKKR{^{0A}_{iA}} \right)
     +
     \dKKR{^0_i}
     \right]
     \Bigg)
    dS_{0j}
    \nn
\\
  &&
   -
   (n-2)
 \int_{\nored{\partial E(R)}}
   \big(
   { g^{\mu[0} - e^{-1}}
    \backg^{\mu[0} \big) \bcov{_\mu} X^{i  ]}
     dS_{0i}
    \Bigg\}
   \,.
    \label{27III17.1+}
\end{eqnarray}
\item
   Suppose instead that $X^a=0$, thus $X$ has only  components tangential to the Kaluza-Klein fibers. Then, again at $x^0=0$,
\begin{eqnarray}
 \nn
  \nored{ \HboundX }
  & = &
   \lim_{R\rightarrow\infty} \Bigg\{
    \frac{3}{16 (n-1)\pi}
   \int_{\partial E(R)}
    X^A
    \nored{Z}^{e} ( \dKKR{_{Ab}^{fb}} )
   dS_{ef}
    \nn
\\
  &&
    -
  \frac{1}{8\pi}
   \int_{\nored{\partial E(R)}}
   \big(
   { g^{\mu[0} - e^{-1}}
    \backg^{\mu[0} \big) \bcov{_\mu} X^{i  ]}
     dS_{0i}
     \Bigg\}
   \,.
    \label{2III17.2+}
\end{eqnarray}
\end{enumerate}

\subsection{$(n\pKKN)+1$--decomposition}
 \label{ss5IV17.1}

In a Cauchy-data context it is convenient to express the global charges explicitly in terms of Cauchy data. Here one can  use the Gauss-Codazzi-Mainardi embedding equations to reexpress our space-time-Riemann-tensor integrals in terms of the Riemann tensor of the initial-data metric and of the extrinsic curvature tensor. For this
we consider $X^\mu=\delta^\mu_0$ and $x^0=0$, i.e., we consider \eqref{17IV17.2+}.

We start with the case of KK asymptotically flat initial data sets. Keeping in mind our convention that $(x^I)=(x^i,x^A)$, we can replace $\KKR{^I_{JKL}}$ with the $(n\pKKN) $-dimensional Riemann tensor, which we denote by  $\spKKR{^I_{JKL}}$,
by means of the Gauss-Codazzi relation
\begin{equation}
 \KKR{^I_{JKL}}=\spKKR{^I_{JKL}}+o(r^{-2\alpha-2})\,.
\end{equation}
Hence, from \eq{17IV17.2} we obtain
\begin{eqnarray}
  p_0
  &=&
   -\frac{1}{8 (n-2) \pi }
   \lim_{R\rightarrow\infty}\int_{S(R)} \int_{\mathbb{T}^N}
    x^j
    \left(\spKKR{^{i}_{j}} + \frac 1 {2(n-1)}\spKKR{^{k}_{k}}\delta^i_j
    +
    \frac{1}{n-1}
    \spKKR{^{iA}_{jA}}
     \right)
      N
    d\spaceS_{i}
    \,.
     \label{23VI17.2}
\end{eqnarray}
We note that in the usual asymptotically flat case, $\KKN=0$, the last integral is not present. Further, $\spKKR{^{k}_{k}}$ becomes then the Ricci scalar of the initial data metric, with $\spKKR{^{k}_{k}}=o(r^{-2\alpha -2})$ because of the scalar constraint equation, and hence does not contribute to the integral. The above reproduces thus the well-known-by-now formula for the ADM mass in terms of the Ricci tensor of the initial data metric
 \cite{MiaoTamMass,HuangCenterOfMass,HerzlichRicciMass,CarlottoSchoen} when the Ricci scalar decays fast enough, as we assumed here.

We pass now to the case covered in Section~\ref{ss23VI17.2}, namely $\KKN =0$ but $\Lambda <0$, with the background metric $\backg$ is as in \eq{24V17.21}-\eq{24V17.22}.  Let  $k_{IJ}$ be the extrinsic curvature tensor of the slices $\{x^0=\const\}$. If we assume that $k_{IJ}$ satisfies
\bel{23VI17.5}
 |k| := \sqrt{  g^{IJ} g^{LM} k_{IL}k_{JM}}= o (r^{-n/2})
 \,,
\ee
from \eq{24V17.222} we obtain a formula first observed in~\cite{HerzlichRicciMass}:
\begin{eqnarray}
  \HboundX
    &= &
   - \frac{1}{16(n-2)\pi}
   \lim_{R\rightarrow\infty}\int_{\boundaryset }  X^0 Z^j
    ( \spKKR{}^i{}_j - \frac {\spKKR{}}{n}\delta^i_j) N
    d\sigma_i
     \,,
           \label{23VI17.6}
\end{eqnarray}
where in \eq{23VI17.6} we have assumed that $X$ is a Killing vector of the anti-de Sitter background which is normal to the hypersurface $\{t=0\}$.

Finally, consider general configurations as in Section~\ref{ss23VI17.1}. Under the hypothesis that
\bel{23VI17.3}
 |k|^2 |Z||\partial E(R)|\to_{R\to\infty}0
 \,,
\ee
from \eqref{17IV17.2+} we find
\begin{eqnarray}
 \nn
  \nored{ \Hb \left(\partial_0,\mcS \right)}
  &=&
   \frac{1}{8 (n-2) \pi  }
   \lim_{R\rightarrow\infty} \Bigg\{
   \int_{\partial E(R)}
   \Bigg(
     \left[
       \frac{1}{2}
        \nored{Z}^i \left( \dKKR{^j_j} \right)
       -
        \nored{Z}^j  \left( \dspKKR{^{i}_{j}} \right)
      \right]
   \\
   \nn
   &&
   -
    \frac{1}{2(n-1)}
      \Bigg[
     \nored{Z}^{i}
     \left(
       \left(\dKKR{^{A}_{A}}\right)
       -
       \left(\dspKKR{^{A}_{A}} \right)
     \right)
       -
       2
       \nored{Z}^{j}
       \left(
         \left( \dspKKR{^{i}_{j}} \right)
         -
         \left(\dspKKR{^{ik}_{jk}}\right)
       \right)
    \Bigg]
    \Bigg)
    N d\sigma_i
    \\
    &&
   \nored{
   -
   (n-2)
    \lim_{R\rightarrow\infty}\int_{\nored{\partial E(R)}}
   \big(
   { g^{\mu[0} - e^{-1}}
    \backg^{\mu[0} \big) \bcov{_\mu} X^{i  ]}
    N
    d\sigma_i
     }
     \Bigg\}
    \,. 
     \phantom{xxxxx}
     \label{23VI17.4}
\end{eqnarray}
%

%
\section{Komar integrals }
  \label{22II17}

If $X^\alpha$ is a Killing vector field of both $g$ and $\backg$, we have
\begin{equation}
 X^\mu \KKR{_{\mu bcd}}
 =
 \nabla_b \nabla_c X_d
 \,,
 \qquad \text{and}
 \qquad
 X^\mu \bKKR{_{\mu bcd}}
 =
 \bcov{_b} \bcov{_c} X_d
 \,.
\end{equation}
This allows us to express some of the integrals above as Komar-type integrals.

We start with the set-up of Section~\ref{ss23VI17.1}; the KK-asymptotically flat case can be obtained directly from the calculations here by setting $\bKKR{_{\alpha\beta\gamma\delta} }= 0$.
To make things clear: we assume \eq{3V17.11}-\eq{5V17.4}, \eq{25VI17.1}-\eq{19IV17.1++}, together with \eq{24V17.31}-\eq{25VI17.11}, and recall that all these hypotheses are satisfied under the corresponding hypotheses made in the KK-asymptotically flat case.

The contribution from the first integrand in \eqref{31III17.5+} can be manipulated as 
\footnote{Strictly speaking, under our asymptotic conditions each individual integrand might fail to have a finite limit as $R\to \infty$, it is only the integral of the sum of all terms which is guaranteed to have a limit. A careful reader will make the calculation below  with the remaining terms from  \eqref{31III17.5+} added to each integral.} 
%
%
\begin{eqnarray}
 \nn
  \lefteqn{
  \lim_{R\rightarrow\infty}\int_{\partial E(R)}
   X^a Z^b \left( \dKKR{_{ab}^{ef}} \right) dS_{ef}
   } &&
    \\
    \nn
    &=&
    \lim_{R\rightarrow\infty}\int_{\partial E(R)}
    \left[
    \left(
     \tensor{X}{^{[f;e]}_{;b}}
     -
     \tensor{X}{^{[f{\|} e]}_{{\|} b}}
    \right)
    Z^b
   -
   X^A Z^b \left( \dKKR{_{Ab}^{ef}} \right)
   \right]
    dS_{ef}
   \\
   \nn
   &=&
     \lim_{R\rightarrow\infty}\int_{\partial E(R)}
    \Bigg\{
      (n-1)
      \left(
        X^{[e;f]}
        -
        X^{[e{\|} f]}
       \right)
      -
      3 \left(
          X^{[e;f} Z^{b]}
          \right)_{;b}
      +
        3 \left(
          X^{[e{\|} f} Z^{b]}
          \right)_{{\|} b}
       \\
       \nn
       &&
       +
       2
       \left(
        \KKR{_{\mu b}^{b [f}} Z^{e]}
        -
        \bKKR{_{\mu b}^{b [f}} Z^{e]}
       \right)
       X^\mu
    dS_{ef}
    -
   X^A Z^b \left( \dKKR{_{Ab}^{ef}} \right)
    \Bigg\}
     dS_{ef}
   \\
   \nn
   &=&
    \lim_{R\rightarrow\infty}
    \Bigg\{
   (n-1)
   \int_{\partial E(R)}
     \left(
        X^{[\alpha;\beta]}
        -
        X^{[\alpha {\|} \beta]}
     \right)
     dS_{\alpha\beta}
     +
      \int_{\partial E(R)}
    \Big[ 2 X^\mu Z^e \left( \dKKR{^{fb}_{b\mu}} \right)
     \\
     &&
     -
     X^A Z^b \left( \dKKR{_{Ab}^{ef}} \right)
     \Big]
    dS_{ef}
    \Bigg\}
    \,,
\end{eqnarray}
where the semicolon ($;$) denotes the covariant derivative of the metric $g$ and the double bar (${\|}$) denotes the covariant derivative of the background metric $\backg$. Moreover, we used the Gauss's theorem, e.g.
\begin{equation}
  \lim_{R\rightarrow\infty}\int_{\partial E(R)}\left(X^{[e{\|} f} Z^{b]}\right)_{{\|} b} \sqrt{|\det g|}
 \, d\Sigma_{ef}=
  \lim_{R\rightarrow\infty}\int_{\partial E(R)}\left(X^{[e{\|} f} Z^{b]}\right)_{{\|} b} \sqrt{|\det \backg|}
 \, d\Sigma_{ef}=0\,.
\end{equation}
Hence, under the hypotheses used in the derivation of \eqref{31III17.5+},  we can rewrite \eqref{31III17.5+}   as
\begin{eqnarray}
 \nn
 \lefteqn{
  \HboundX
 =
  \frac{1}{16  (n-2) \pi }
    \lim_{R\rightarrow\infty} \Bigg\{
    \int_{\partial E(R)}
    (n-1)
    \left(
        X^{[\alpha;\beta]}
        -
        X^{[\alpha {\|} \beta]}
     \right)
    dS_{\alpha\beta}
}
&&
    \\
    \nn
    &&
   +
    \int_{\partial E(R)}
    \Bigg(
     2 X^\mu Z^e \left( \dKKR{^{fb}_{b\mu}} \right)
     -
     X^A Z^b \left( \dKKR{_{Ab}^{ef}} \right)
   +
     4 Z^{[e}X^{a]} \left( \dKKR{^f_a} \right)
     \\
     \nn
     &&
     -
     Z^e X^f \left( \dKKR{^c_c} \right)
   -
    \frac{1}{n-1}
   \Big[
    (n-3) Z^e X^f
     (\dKKR{^{bA}_{bA}})
    +
     4 Z^{[e} X^{a]} \left( \dKKR{^{fA}_{aA}} \right)
    \\
    &&
    +
   3 (n-2)
    X^A
    Z^{e}
     \left( \dKKR{^{fb}_{b A}} \right)
      \Big]
      \Bigg)
   dS_{ef}
-
    (n-2)
   \lim_{R\rightarrow\infty}\int_{\nored{\partial E(R)}}
    \big(
   { g^{\mu[a} - e^{-1}}
    \backg^{\mu[a} \big) \bcov{_\mu} X^{b  ]}
      dS_{ab}
      \Bigg\}
   \,.
\end{eqnarray}
The first integrand is the difference of Komar integrands of $g$ and $\backg$.

Specialising to the KK-asymptotically flat case for background-covariantly constant Killing vectors, this reads
\begin{eqnarray}
 \nn
   p_\mu X^\mu_\infty
  & = &
  \frac{1}{16  (n-2) \pi }
   \lim_{R\rightarrow\infty} \Bigg\{
   (n-1)
   \int_{S(R)} \int_{\mathbb{T}^N}
    X^{\alpha;\beta} dS_{\alpha\beta}
   \\
   \nn
   &&
   +
   \int_{S(R)} \int_{\mathbb{T}^N}
    \Bigg(
     2 X^\mu x^e \KKR{^{fb}_{b\mu}}
     -
     X^A x^b \KKR{_{Ab}^{ef}}
   -
    \frac{1}{n-1}
   \Big[
    (n-3) x^e X^f
     \KKR{^{bA}_{bA}}
    +
     4 x^{[e} X^{a]} \KKR{^{fA}_{aA}}
    \\
    &&
    +
    3(n-2)
    X^A
    x^{e} \KKR{^{fb}_{b A}}
    \Big]
    \Bigg)
   dS_{ef}
   \Bigg\}
   \,.
\end{eqnarray}

It appears thus that in general  Komar-type integrals do \emph{not} coincide with the Hamiltonian generators.
This is really the case, as can be seen for the Rasheed solutions. Using \eq{31III17.11} one readily finds for $X=\partial_t$, keeping in mind that $n=3$:
%
\bel{22X17}
 \frac{1}{8\pi  }
   \lim_{R\rightarrow\infty}\int_{S(R)} \int_{S^1}
    X^{\alpha;\beta} dS_{\alpha\beta}
     =
       \left\{
         \begin{array}{ll}
           2 \pi \big(M+ \frac{\Sigma}{\sqrt 3}
      \big), & \hbox{$P=0$;} \\
          8  \pi P\big(M+ \frac{\Sigma}{\sqrt 3}
      \big), & \hbox{$P\ne 0$,}
         \end{array}
       \right.
\ee
which does \emph{neither} coincide with $p_0$, cf.~\eq{31III17.12}, \emph{nor} with the ADM mass of the space metric $g_{ij}dx^i dx^j$.  Note that the Komar integral of the space-time metric $g_{ab}dx^a dx^b$ will equal   $M+ \frac{\Sigma}{\sqrt 3}  $ regardless of the value of $P$.

Next,  for $X=\partial_4$ we obtain
\bel{22X17+}
 \frac{1}{8\pi  }
   \lim_{R\rightarrow\infty}\int_{S(R)} \int_{S^1}
    X^{\alpha;\beta} dS_{\alpha\beta}
     =
       \left\{
         \begin{array}{ll}
          4\pi Q, & \hbox{$P=0$;} \\
          16\pi P Q, & \hbox{$P\ne 0$,}
         \end{array}
       \right.
\ee
which  is twice  the Hamiltonian charge $p_4$.

As a simple application of \eq{22X17}, suppose that there exists a Rasheed metric without a black hole region. Since the divergence of the Komar integrand is zero, we obtain $M = -\Sigma/\sqrt 3$. But this is precisely one of the parameter values excluded in the Rasheed metrics, cf.\ \eq{16VI10.4} below.
We conclude that the regular metrics in the Rasheed family must be black-hole solutions.

For the case of metrics which asymptote to a maximally symmetric background $\backg$ with  $\Lambda \ne 0$, as in Section~\ref{ss23VI17.2}, the Komar integral resulting from  \eqref{24V17.222} reads
\begin{eqnarray}
 \HboundX
  &=&
  \frac{1}{16(n-2)\pi}
   \lim_{R\rightarrow\infty}\int_{\boundaryset }  X^\nu Z^\xi
     \noKKW{^{\alpha\beta}_{\nu\xi}}
     dS_{\alpha\beta}
     \nn
     \\
     &=&
     \lim_{R\rightarrow\infty}
     \Bigg\{
     \frac{n-1}{16(n-2)\pi}
     \int_{\boundaryset }  X^{[\alpha;\beta]}
     dS_{\alpha\beta}
     -
     \frac{\Lambda}{4 (n-2)(n-1)n\pi}
     \int_{\boundaryset } \,  X^\alpha Z^\beta dS_{\alpha\beta}
     \Bigg\}
     \,.
\end{eqnarray}
%

\section{Witten's positivity argument}
  \label{s14II17.5}

The Witten positive-energy argument~\cite{Witten:mass,Bartnik} (compare~\cite{DeserTeitelboim})  generalises in an obvious manner to KK-asymptotically flat metrics.
Assuming that the initial data hypersurface $\hyp$ is spin, we consider the Witten boundary integral $\Witten$ defined as
\beal{spinmass}
 & \displaystyle
 \Witten (\phi_\infty) := \lim_{R\to\infty} \int_{S(R)\times \T^\KKN } \mcU^i
  d\spaceS_i\,,&
\\
 &\mcU^I=\langle \phi, D^I \phi+ \ga^I\!\! \mtd\phi\rangle
 \,,
\eeal{spinmass2}
where 
$\phi$
is a spinor field which asymptotes to a constant spinor $\phi_\infty$ at an
appropriate rate as one recedes to infinity in the asymptotic end, and $\mtd:=\ga^J D_J$ is the Dirac operator on $\hyp$. (Note that the asymptotic spinors $\phi_\infty$ might be incompatible with the spin structure of $\hyp$, in which case the argument below does of course not apply; compare~\cite{BrillPfister,WittenKK,WittenBanff}.) It is standard to show that in the natural spin frame we have
\bea
 \mcU^I&=&\frac
    14\sum_{L=1}^{n\pKKN }(\partial_L g_{IL}-\partial_I g_{LL})
    |\phi_\infty|^2+o(r^{-2\alpha-1})\,.
\eeal{Uagain4}
Assuming positive and suitably decaying energy density on a maximal (i.e., $g^{IJ}K_{IJ}=0$) initial data hypersurface,
 such that
\bel{21IV17.1}
 \mbox{$\hyp$ is metrically complete and either is boundaryless or has a trapped compact boundary,}
\ee
the proof of existence of the desired solutions of the Witten equation $\mtd \phi=0$ can be carried-out along lines identical to the usual asymptotically flat case, cf.\ e.g. \cite{BartnikChrusciel2,Herzlich:mass}.
Comparing with \eq{a.12++}, we conclude that positivity of $\Witten$ is equivalent to positivity of the Hamiltonian mass:
$$
 p_0\ge 0
 \,.
$$

It should be emphasised that $p_0$ does \emph{not} necessarily coincide with the ADM mass of $g_{IJ} dx^I dx^J$.

The above argument required positivity of the scalar curvature of $g_{IJ}dx^I dx^J$. This is not needed if one replaces in \eq{spinmass2} the usual spinor covariant derivative by
 \bel{Sli6}
 D_I \to D_I + \frac 12 K_I{^J}\ga_J\ga_0 \,.
 \ee
The Witten quadratic form $\mycal W$ becomes instead
\begin{eqnarray}
 \label{Sli9}
    \lim_{R\to\infty}\oint_{S(R)\times \T^\KKN }\mcU^i
 d\spaceS_i = {4\pi} p_\mu\langle
 \phi_\infty,\ga^\mu\ga^0\phi_\infty\rangle\
  \,,
\end{eqnarray}
and is non-negative for all $\phi_\infty$ when the dominant energy condition is assumed on initial data hypersurfaces as in \eq{21IV17.1}.
The positivity of $\mycal W$ is equivalent to timelikeness of the $(n\pKKN +1)$-vector $p_\mu$. Equivalently,
\bel{20IV17.6}
p_0^2 - \sum_{i=1}^n p_i^2 \ge \sum_{A=n+1}^{n\pKKN } p_A^2
  \ge 0
 \,.
\ee
The first inequality is saturated if and only if the initial data set can be isometrically embedded in $\R\times \R^n \times \T^\KKN $ equipped with the flat Lorentzian metric (compare~\cite{ChMaerten}).

 As an example, consider the Rasheed metrics with $P=0$. The corresponding domains of outer communications have topology $\R\times S^1 \times (\R^3 \setminus B(R))$, where the $\R$ factor corresponds to the time variable,   $S^1$ is  the Kaluza-Klein factor, and the $\R^3 \setminus B(R)$ factor describes the space-topology of the black hole. It thus has the obvious spin structure inherited from a flat $\R\times S^1 \times  \R^3$, together with the obvious associated parallel spinors. Therefore the Witten-type argument just described applies, leading to
\bel{20IV17.7}
 M^2 \ge  {Q^2}
 \,,
\ee
with the inequality strict for black-hole solutions.
If we denote by $M_\spADM$ the  ADM mass of the three-dimensional-space part of the Rasheed metric, this can equivalently be rewritten as
\bel{20IV17.7+}
 \big(M_\spADM+\frac \Sigma {\sqrt 3}\big)^2 \ge  {Q^2}
 \,,
\ee
compare~\cite{GibbonsWells}.

Note that \eq{20IV17.7+} does not exclude the possibility of negative or vanishing $M_\spADM$ (compare~\cite{HorowitzWiseman,BrillPfister,lebrun-cex}).
We have not attempted a systematic analysis of this issue, and only checked that all Rasheed solutions with $a=0$ and $M=0$ have naked singularities outside of the horizon.

\section{Summary}
 \label{s23IX17.1} 

In this work we have considered families of metrics asymptotic to various background metrics, and studied the Hamiltonians associated with the flow of Killing vectors of the background. We have derived several new formulae for these Hamiltonians, generalising previous work by allowing a cosmological constant, or non-standard backgrounds, and allowing higher dimensions.  In particular:

We have derived an ADM-type formula for  Hamiltonians generating time translations for a wide class of background metrics, cf.\   \eq{23IV17.1+}.

We have provided a formula  for Hamiltonians generating translations for KK-asymptotically flat metrics in terms of the space-time curvature tensor, \Eq{31III17.5}.

We have derived a formula for Hamiltonians associated with generators of all background Killing fields for asymptotically anti-de Sitter space-times  in terms of the space-time curvature tensor, \Eq{24V17.222}.

\Eq{31III17.5+} provides a similar formula for a wide class of backgrounds with $\Lambda=0$.

Equations~\eq{23VI17.6} and \eq{23VI17.4} provide space-and-time decomposed versions of the last two Hamiltonians.

In Section~\ref{22II17} we have derived several Komar-type formulae for the Hamiltonians above for vector fields $X$ which are Killing vectors for both the background and the physical metric.

In Section~\ref{s14II17.5} we have pointed-out the consequences of a Witten-type positivity argument for $KK$-asymptotically flat space-times: instead of proving positivity of the ADM energy, the argument provides an inequality involving the Kaluza-Klein charges and the energy. An explicit version of the inequality has been established for KK-asymptotically flat Rasheed metrics.

In addition to the above, we have carried-out a careful study of Rasheed metrics, Appendix~\ref{s14II17.6} below, to obtain a non-trivial family of metrics with singularity-free domains of outer communications to which our formulae apply. We have pointed out the restrictions \eq{16VIII20.2} and \eq{16VIII20.2asdf} on the parameters needed to guarantee absence of naked singularities in the metric. We have shown that all metrics satisfying these conditions together with $P=0$ have a stably causal domain of outer communications, and we have given sufficient conditions for stable causality when $P \ne 0$ in \eq{23IX17.11}.  In Appendix~\ref{s24IX17.1} we have pointed out that the Rasheed metrics with $P\ne 0$ are not $KK$-asymptotically flat, and described their asymptotics. We have determined their global charges, which are significantly different according to whether or not $P$ vanishes.

Last but perhaps not least, \eq{2III17.4} provides a useful identity, which we haven't met in the literature, satisfied by the Riemann tensor in any dimensions and generalising the usual double-dual identity valid in four dimensions.   

\appendix

 \section{An example: Rasheed's solutions}
  \label{s14II17.6}
  \newcommand{\fsquare}{\nored{G}}

D.~Rasheed~\cite{Rasheed} has constructed a family of stationary axi-symmetric {solutions of the  five-dimensional   vacuum Einstein} equations which take the form
\be
 ds_{(5)}^2
	=
	{B\over A}\left(d\nored{x^4} + 2A_\mu dx^\mu\right)^2 + \sqrt{A\over
	B}ds_{(4)}^2
\,,
\label{16VI10.1}
\ee
where $a$, $M$, $P$, $Q$ and $\Sigma$ are real numbers satisfying
\bea
 &
 \frac{Q^2}{\Sigma + M\sqrt{3}}+\frac{P^2}{\Sigma - M\sqrt{3}}
	    =
		\frac{2 \Sigma}{3}
    \,,
     &
 \label{16VI10.22}
\\
%
%
 &
   M^2+\Sigma^2-P^2-Q^2 \ne 0
   \,,
   \quad
    \left(M+\Sigma/\sqrt{3}\right)^2-Q^2 \ne 0
   \,,
   \quad
    \left(M-\Sigma/\sqrt{3}\right)^2-P^2 \ne 0
    \,,
     \label{11VI17.2}
     &
\\
 &
 M \pm \frac \Sigma {\sqrt{3}} \ne 0
  \,,
  \quad
 \nored{F^2}
	 :=
	{\left[\left(M+\Sigma/\sqrt{3}\right)^2-Q^2\right]
	\left[\left(M-\Sigma/\sqrt{3}\right)^2-P^2\right]\over M^2+\Sigma^2-P^2-Q^2} > 0
\;,
 &
\label{16VI10.4}
\eena
and where
\be
 ds_{(4)}^2
	=
	-{\fsquare \over\sqrt{AB}}\left(dt+{\omega^0}_\phi d\phi\right)^2 +
	{\sqrt{AB}\over\Delta}dr^2 + \sqrt{AB}d\nored{\theta}^2 + {\Delta\sqrt{AB}\over
	\fsquare }\sin^2(\nored{\theta}) d\phi^2
\,, \label{16VI10.2}
\ee
with
%
\bena
 A
	&=&
	\left(r-\Sigma /\sqrt{3}\right)^2 - {2P^2\Sigma\over\Sigma - M\sqrt{3}} +
	a^2\cos^2(\nored{\theta}) + {2JPQ\cos(\theta)\over\left(M+\Sigma/\sqrt{3}\right)^2-Q^2}
\;,
\nn
\\
\nn
\\
 B
	&=&
	\left(r+\Sigma /\sqrt{3}\right)^2 - {2Q^2\Sigma\over\Sigma + M\sqrt{3}} +
	a^2\cos^2(\nored{\theta}) - {2JPQ\cos(\theta)\over\left(M-\Sigma/\sqrt{3}\right)^2-P^2}
\;,
\nn
\\
\nn
\\
 \fsquare
	&=&
	r^2 - 2Mr + P^2 + Q^2 - \Sigma^2 + a^2\cos^2(\nored{\theta})
\;,
\nn
\\
\nn
\\
 \Delta
	&=&
	r^2 - 2Mr + P^2 + Q^2 - \Sigma^2 + a^2
\;,
\nn
\\
\nn
\\
 {\omega^0}_\phi
	&=&
	{2J\sin^2(\nored{\theta})\over \fsquare }\left[r+E \right]
\;,
\nn
\\
\nn
\\
 J^2
	&=&
	a^2 \nored{F^2}
\;,
\label{16VI10.3}
\eena
whereas $E$ is given by
\bena
 E
	&=&
	 -M+{\left(M^2+\Sigma^2-P^2-Q^2\right)\left(M+\Sigma
	/\sqrt{3}\right)\over\left(M+\Sigma /\sqrt{3}\right)^2-Q^2}
\;.
\label{16VI10.4+}
\eena
The  physical-space Maxwell potential is given by
\bena
 2A_\mu dx^\mu
	=
	{C\over B}dt + \left({\omega^5}_\phi + {C\over
	B}{\omega^0}_\phi\right)d\phi
\,,
\label{16VI10.5}
\eena
where
\bena
 C
	&=& 2Q\left(r-\Sigma /\sqrt{3}\right) - {2PJ\cos(\theta)\left(M+\Sigma
	/\sqrt{3}\right)\over\left(M-\Sigma /\sqrt{3}\right)^2-P^2}
\;,
\label{16VI10.6}
\\
 {\omega^5}_\phi
 	&=&
	\frac{H}{\fsquare }\;,
\label{16VI10.7}
\eena
and
\be
 H
	:=
	{2P\Delta}\cos(\theta) - {2QJ\sin^2(\nored{\theta})
	\left[r\left(M -
	\Sigma/\sqrt{3}\right) + M\Sigma/\sqrt{3} + \Sigma^2-P^2-Q^2\right] \over
	\left[\left(M+\Sigma/\sqrt{3}\right)^2-Q^2\right]}
\,.
\label{16VI10.8}
\ee

The Rasheed metrics \eq{16VI10.1}
 have been obtained by applying a solution-generating technique (\cite{Rasheed}, compare~\cite{CGerard})
to the Kerr metrics. This guarantees that these metrics solve the five-dimensional vacuum Einstein equations when the constraint  \eq{11VI17.2} is satisfied. As the procedure is somewhat involved, it appears useful to crosscheck the vanishing of the Ricci tensor using computer algebra. We have been able to verify this in the $P=0$ case with {\sc Sage} (which required a week-long computation on a personal computer),
as well as for a set of samples for the parameters $(M,\,a,\,P,\,Q,\,\Sigma)$
in the $P \neq 0$ case with {\sc Mathematica}. We have, however, not been able to do it for the full set of parameters.

Let us address the question of the global structure  of the metrics above. We have
$$
 \det g = -A^2 \sin^2(\theta)
\,,
$$
which shows that the metrics  are smooth and Lorentzian except possibly at the zeros of $A$, $B$, $G$, $\Delta$, and $\sin(\theta)$.

 After a suitable periodicity of $\phi$ as in Section~\ref{s24IX17.1} below has been imposed,  regularity at the axes of rotation away from the zeros of denominators follows from the factorisations
\beal{24IV17.1}
  \left(\frac{\Delta}{G}-1 \right)
   &   =
 &         \frac{a^2 \sin ^2(\nored{\theta} )}{a^2 \cos ^2(\nored{\theta} )-2 M r+P^2+Q^2+r^2-\Sigma ^2}
\,,
\\
    2 A_\phi
     - 2 P \frac{\Delta}{G} \cos(\nored{\theta})
    &=&
   \frac{\sin ^2 (\nored{\theta})}{\fsquare}
   \left( \mathcal{H} +{2J C \over  B }\left[r+E \right]  \right)
    \,,
\eea
where
\be
 \mathcal{H}
	 :=
	  -{2QJ
	\left[r\left(M -
	\Sigma/\sqrt{3}\right) + M\Sigma/\sqrt{3} + \Sigma^2-P^2-Q^2\right] \over
	\left[\left(M+\Sigma/\sqrt{3}\right)^2-Q^2\right]}
\,.
\ee

It will be seen below  that, after restricting the parameter ranges as in \eq{16VIII20.2} and \eq{16VIII20.2asdf}, the location of Killing horizons is determined by the zeros of
\bena
	\Bigg |
	\begin{array}{ccc}
	g_{tt} & g_{t \phi} & g_{t 4} \\
	g_{\phi t} & g_{\phi \phi} & g_{\phi 4} \\
	g_{4 t} & g_{4 \phi} & g_{4 4} \\
	\end{array}
	\Bigg | =
	       - \Delta \sin^2  (\theta)
\,,
\eena
and thus by the real roots  $r_+ \geq r_-$  of $\Delta$, if any:
\bena%
 r_\pm
	=
	M\pm\sqrt{M^2+\Sigma^2-P^2-Q^2-a^2}
\,.
\label{16VIII20.3}%
\eena
%

\subsection{Zeros of the denominators}
{
The norms
$$
 g_{tt} = \frac {W} { AB}
  \ \mbox{and} \
 g_{44} =   \frac B {A}
\,,
$$
of the Killing vectors $\partial_t$ and $\partial_4$ are geometric invariants, where $W=-GA+C^2$.
}
So zeros of  $A$ and of $A B$ correspond to singularities in the five-dimensional geometry except
if
\begin{enumerate}
\item  a zero of $A$ is a joint zero of  $A$, $B$ and $W$, or if
\item
 a zero of $B$ which is not a zero of $A$  is also a zero of $W$.
\end{enumerate}

Setting
\bena
 \mathcal{A} := {2JPQ \over {a^2 \left( \left(M+\Sigma/\sqrt{3}\right)^2-Q^2 \right)}}
\,,
\label{16VIII20.12}
\eena
one checks that if
\bena
 &&
 \left\{
   \begin{array}{ll}
 {2P^2\Sigma\over\Sigma - M\sqrt{3}} - a^2 (1- |\mathcal{A}|)=0, ~~\text{when}~~ \, |\mathcal{A}|> 2 & \hbox{or} \\
 {2P^2\Sigma\over\Sigma - M\sqrt{3}} +  \frac{ a^2 \mathcal{A}^2}{4}=0, ~~ \text{when} ~~ |\mathcal{A}|\leq 2 \,, &
   \end{array}
 \right.
  \label{24X17.2dz}
\eena
then $A$ vanishes exactly at one point. Otherwise the set of zeros of $A$ forms a curve in the $(r,\theta)$ plane.
Let  $\theta\mapsto r_A^+(\theta)$ denote the curve, say $\gamma$, corresponding to the set of largest zeros of $A$.

Note that $W$ and $A$ are polynomials in $r$, with $A$ of second order. If $W/A$  is smooth, the remainder of the polynomial  division of $W$ by  $r-r_A^+$ must vanish on the part of $\gamma$ that lies outside the horizon. One can calculate this remainder with {\sc Mathematica}, obtaining a function of $\theta$ which vanishes at most at isolated points, if at all. It follows that the  division of $W$ by $A$  is singular on the closure of the domain of outer communications (d.o.c.), i.e.\ the region $\{r \ge r_+\}$, if $A$ has zeros there, except perhaps when \eqref{24X17.2dz} holds.

One can likewise exclude  a joint zero of $W$ and $B$ in the closure of the d.o.c.\ without a zero of $A$, except possibly for the case where this zero is isolated for $B$ as well, which happens if
\bena
 &&
 \left\{
   \begin{array}{ll}
 {2Q^2\Sigma\over\Sigma + M\sqrt{3}} - a^2 (1- |\mathcal{B}|)=0, ~~\text{if}~~ \, |\mathcal{B}| > 2 & \hbox{or} \\
 {2Q^2\Sigma\over\Sigma + M\sqrt{3}} +  \frac{ a^2 \mathcal{B}^2}{4}=0, ~~ \text{if} ~~ |\mathcal{B}|\leq 2 \,. &
   \end{array}
 \right.
  \label{24X17.1dz}
\eena
See \cite{HoerzingerRasheed} for a more detailed analysis of the borderline cases.

Summarising: a necessary condition for a black hole without obvious singularities in the closure of the domain of outer communications is that all zeros of $A$ lie under the outermost Killing  horizon $r=r_+$.
One finds that this will be the case if and only if
\bena
 &&|\mathcal{A}|>2 ~~\text{and}~~
 \left\{
   \begin{array}{ll}
 {2P^2\Sigma\over\Sigma - M\sqrt{3}} - a^2 (1- |\mathcal{A}|)<0, & \hbox{or} \\
 M+\sqrt{M^2+\Sigma^2-P^2-Q^2 -a^2} > \frac{\Sigma}{3}
 +\sqrt{
 {2P^2\Sigma\over\Sigma - M\sqrt{3}} - a^2 (1- |\mathcal{A}|)
 }, & \hbox{}
   \end{array}
 \right.
\nn
\\
\text{or}
\nn
\\
 &&|\mathcal{A}| \leq 2 ~~\text{and}~~
 \left\{
   \begin{array}{ll}
 {2P^2\Sigma\over\Sigma - M\sqrt{3}} +  \frac{ a^2 \mathcal{A}^2}{4}<0, & \hbox{or} \\
 M+\sqrt{M^2+\Sigma^2-P^2-Q^2 -a^2} > \frac{\Sigma}{3}+
 \sqrt{
 {2P^2\Sigma\over\Sigma - M\sqrt{3}} +  \frac{ a^2 \mathcal{A}^2}{4}
}
, & \hbox{}
   \end{array}
 \right.
\label{16VIII20.2}
\eena
except perhaps when \eqref{24X17.2dz} holds.

An identical argument applies to the zeros of $B$, with the zeros of $B$ lying on a curve unless \eqref{24X17.1dz} holds.
Ignoring this last case, the zeros of $B$ need similarly be hidden behind the outermost Killing horizon. Setting
\bena
 \mathcal{B} := -{2JPQ \over {a^2 \left( \left(M-\Sigma/\sqrt{3}\right)^2-P^2 \right)}}
\,,
\label{16VIII20.12+}
\eena
one finds that this will be the case if and only if
\bena
 &&|\mathcal{B}|>2 ~~\text{and}~~
 \left\{
   \begin{array}{ll}
 {2Q^2\Sigma\over\Sigma + M\sqrt{3}} - a^2 (1- |\mathcal{B}|)<0, & \hbox{or} \\
 M+\sqrt{M^2+\Sigma^2-P^2-Q^2 -a^2} > - \frac{\Sigma}{3}
 +\sqrt{
 {2Q^2\Sigma\over\Sigma + M\sqrt{3}} - a^2 (1- |\mathcal{B}|)
 }, & \hbox{}
   \end{array}
 \right.
\nn
\\
\text{or}
\nn
\\
 &&|\mathcal{B}| \leq 2 ~~\text{and}~~
 \left\{
   \begin{array}{ll}
 {2Q^2\Sigma\over\Sigma + M\sqrt{3}} +  \frac{ a^2 \mathcal{B}^2}{4}<0, & \hbox{or} \\
 M+\sqrt{M^2+\Sigma^2-P^2-Q^2 -a^2} > - \frac{\Sigma}{3}+
 \sqrt{
 {2Q^2\Sigma\over\Sigma + M\sqrt{3}} +  \frac{ a^2 \mathcal{B}^2}{4}
}
, & \hbox{}
   \end{array}
 \right.
\label{16VIII20.2asdf}
\eena
except perhaps when \eqref{24X17.1dz} holds.

While the above guarantees lack of obvious singularities in the \emph{domain of outer communications} $\{r>r_+\}$ (d.o.c.), there could still be causality violations there. Ideally the d.o.c.\ should be globally hyperbolic, a question which we have not attempted to address. Barring global hyperbolicity, a decent d.o.c. should at least admit a time function, and the function $t$ provides an obvious candidate. In order to study the issue we note the identity
\bena
g^{00}
     &=&
        \frac{
         4 J^2 [r+E]^2\sin^2 (\theta)   - A B  \Delta      }{A \Delta\fsquare}
\,.
\label{17VII22.1}
\eena
A {\sc Mathematica} calculation shows that the numerator factorises through $G$, so that $g^{00}$ extends smoothly through the ergosphere.
When $P=0$ one can verify that $g^{00}$ is negative on the d.o.c. For $P\ne 0$ one can find open sets of parameters which guarantee that $g^{00}$ is strictly negative for $r>r_+$ when $A$ and $B$ have no zeros there.
{
An example is given by the condition
\bel{23IX17.11}
 r_+
     \geq
        \frac{EM+q}{M+E}
\,,
\ee
which is sufficient but not necessary, where $q:=P^2 + Q^2 - \Sigma^2 + a^2$.
}
We hope to return to the question of causality violations in the future.

\begin{figure}[th]
{\includegraphics[scale=.6]{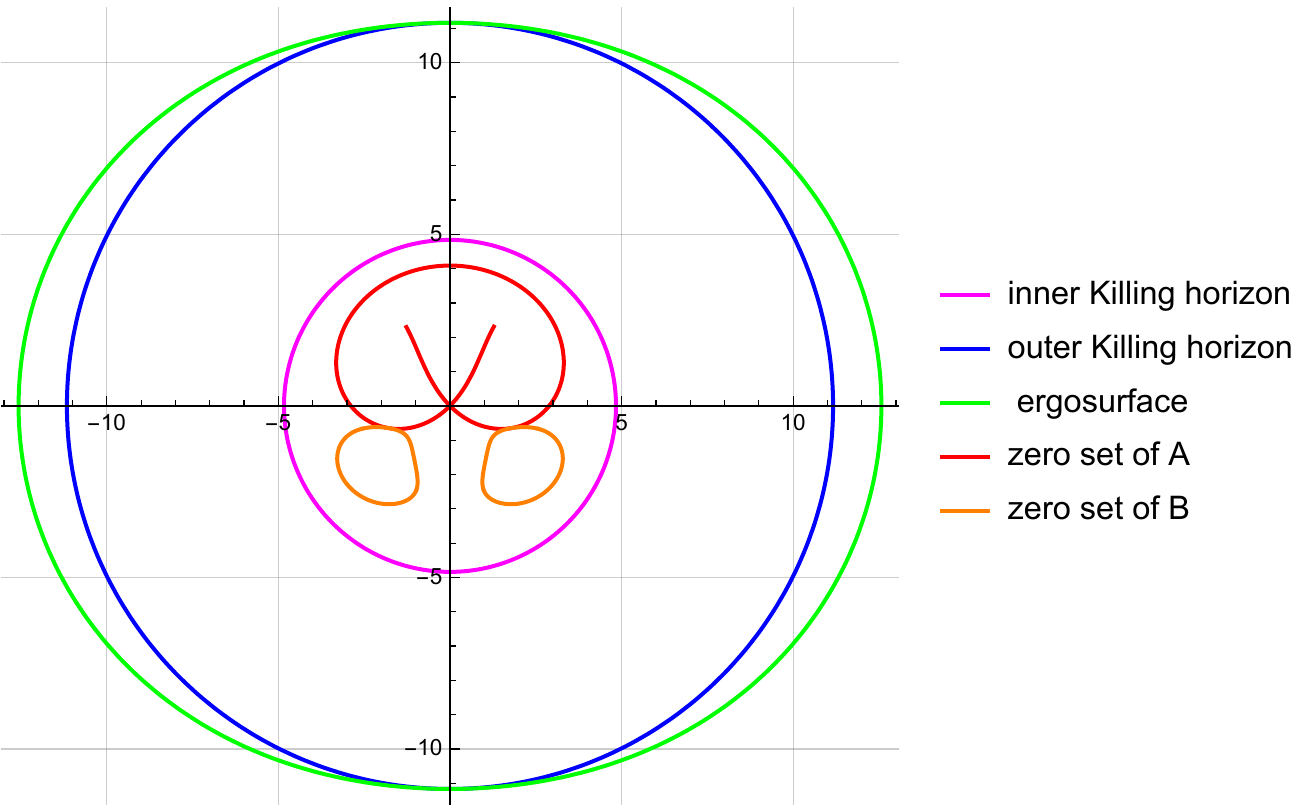}}{\includegraphics[scale=.4]{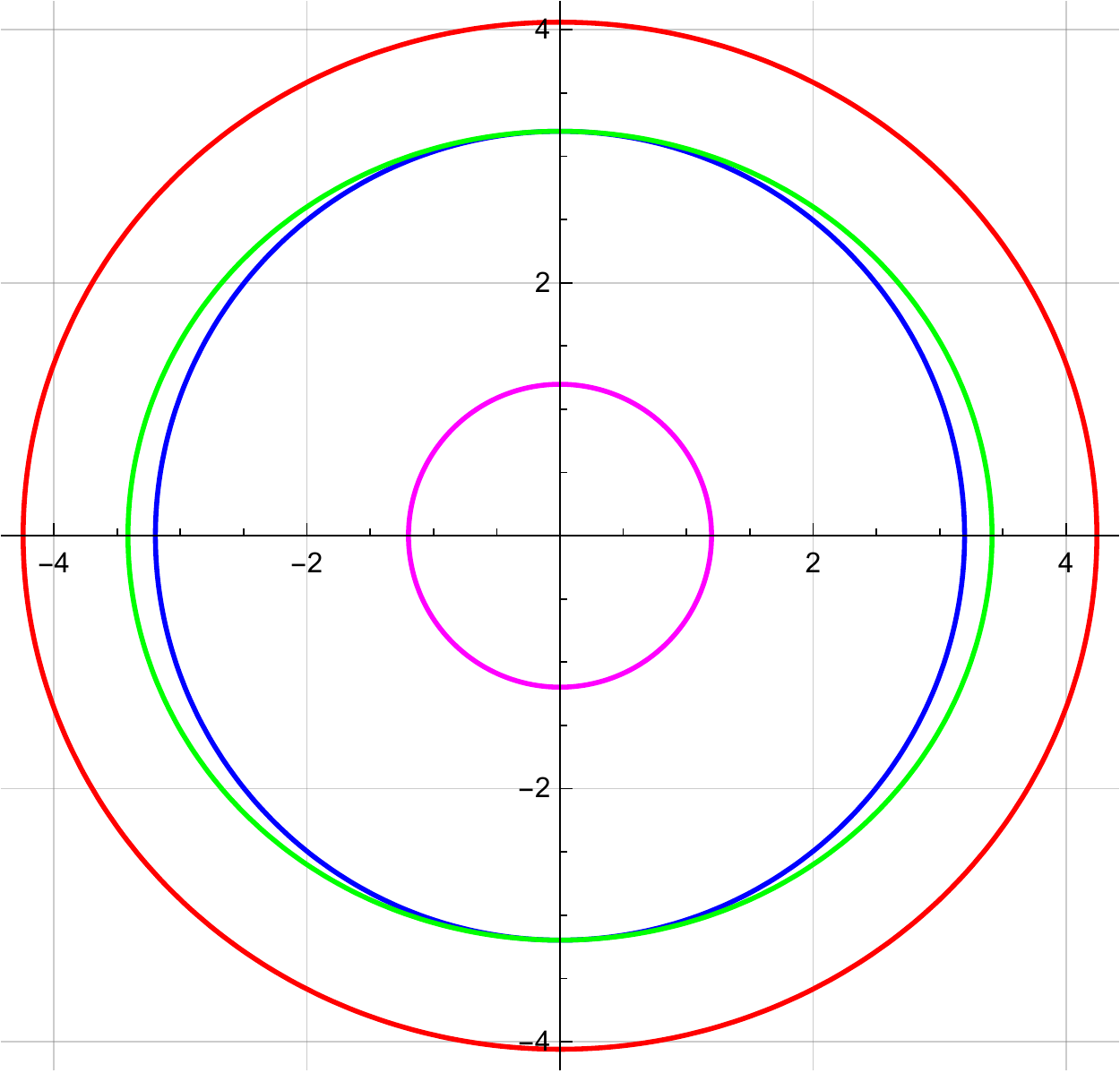}}
%
\caption{Two sample plots for the location of the ergosurface (zeros of $\fsquare$), the outer and inner Killing horizons (zeros of $\Delta$), {and the zeros  of $A\,,B$. Left plot:  $M=8,\,a=\frac{33}{10},\,Q=\frac{8}{5},\,\Sigma = -\frac{23}{5},\,P=-\frac{1}{5} \sqrt{\frac{2 \left(4105960 \sqrt{3}+2770943\right)}{12813}} \approx -7.86$, with zeros of $A$ and $B$ under both horizons, consistently with \eq{16VIII20.2} and \eq{16VIII20.2asdf}. Right plot: $M=1,\,a=1,\,Q=0,\,\Sigma =\sqrt{6},\,P=\sqrt{4-2 \sqrt{2}} \approx 1.08$; here \eq{16VIII20.2} is violated, while the zeros of $B$ occur  at negative $r$.}
\label{Ftobedone}
}
\end{figure}

In Figure~\ref{Ftobedone} we show the locations of the zeros of $A$ and $B$ for some specific sets of parameters satisfying, or violating, the conditions above.

Another potential source of singularities of the metric \eq{16VI10.1} could be the zeros of $G$. It turns out that there are { irrelevant}, which can be seen as follows: The relevant metric coefficient is $g_{\phi\phi}$, which reads
\bena
 g _{\phi \phi}
	=
	{B\over A}   \left({\omega^5}_\phi + {C\over
	B}{\omega^0}_\phi\right)^2  + \sqrt{A\over B} \left( -{\fsquare \over\sqrt{AB}}\left({\omega^0}_\phi \right)^2 +{\Delta\sqrt{AB}\over \fsquare }\sin^2 \theta
\right)
\,.
\label{16VI10.16}
\eena
Taking into account a $G^{-1}$ factor in ${\omega^0}_\phi$, it follows that $g_{\phi\phi}$ can be written as a fraction $(\ldots)/ABG^2$.
 A {\sc Mathematica} calculation shows that the denominator $(\ldots)$  factorises  through $AG^2$, which shows indeed that the zeros of $G$ are innocuous for the problem at hand.

Let us write $ds^2_{(4)}$ as $^{(4)}g_{ab}dx^a dx^b$. The factorisation just described works for $g_{\phi\phi}$ but  does \emph{not} work for $^{(4)}g_{\phi\phi}$. From what has been said we see that the quotient metric $^{(4)}g_{ab}dx^a dx^b$ is always singular in the d.o.c., a fact which  { seems to have been
ignored, and unnoticed,}  in the literature so far.

 \subsection{Regularity at the outer Killing horizon $\mathcal{H}_+$}
 \label{s11VIII17.1}

The location of the outer Killing horizon  $\mathcal{H}_+$ of the Killing field
\bena
	k
	  =
		\partial_t + \Omega_\phi \partial_\phi + \Omega_4 \partial_{x^4}
\,,
\label{17V20.2}
\eena
is given by the larger root $r_+$ of $\Delta$, cf.\  (\ref{16VIII20.3}). The condition that ${\mathcal H}_+$ is a Killing horizon for $k$ is that the  pullback of $
	g_{\mu \nu } k^\nu$
to $\mathcal{H}_+$ vanishes.
This, together with
{
\bena
	\Delta|_{\mathcal{H}_+}=0
\,,
\quad
	\fsquare|_{\mathcal{H}_+}=-a^2 \sin^2 (\theta)
\,,
\eena
}
yields
\bena
\Omega_\phi &=&
				-\frac{1}{{\omega^0}_\phi} \Big{|}_{\mathcal{H}_+} =
           		\frac{a^2}{2 J} (r_+ + E)^{-1}
\,,
\nn
\\
	\Omega_4 &=&
				-\frac{2 (A_t {{\omega^0}_\phi}-{A_\phi })}{{\omega^0}_\phi} \Big{|}_{\mathcal{H}_+} =
             	\frac{Q \left(-3 M r_+-\sqrt{3} M \Sigma +3 P^2+3 Q^2+\sqrt{3} r \Sigma -3 \Sigma ^2\right)}{(E+r_+) \left(3 M^2+2 \sqrt{3} M \Sigma -3 Q^2+\Sigma ^2\right)}
\,.
\label{17V20.1}
\eena
After the coordinate transformation
\bena
	\bar \phi = \phi  - \Omega_\phi \, dt
\,,
\quad
   	\bar  x^4 = x^4  -  \Omega_{4} \, dt
\,,
\eena
the metric (\ref{16VI10.1}) becomes
\bena
g= g_S + \frac{dr^2}{\Delta} + \Delta U dt^2
\,,
\label{17VI21.4}
\eena
where $g_S$ is a smooth $(0,2)$-tensor, with $U:= {g_{tt}}/{\Delta}$ extending smoothly across $\Delta=0$.
Introducing a new time coordinate by
\bena
	\tau =  t  - \sigma \ln(r -r_+)
\,\Rightarrow
\quad
   	d \tau = d t  - \frac{\sigma}{r -r_+} \, dr
\,,
\eena
where $\sigma$ is a constant to be determined,  (\ref{17VI21.4}) takes the form
\bena
g
&=& g_S + \Delta U \left(d \tau + \frac{\sigma}{r-r_+} dr \right)^2 + \frac{dr^2}{\Delta}
\nn
\\
&=& g_S + \Delta U d \tau^2 + \frac{2 \Delta U \sigma}{r - r_+} d \tau dr + \left( \frac{1}{\Delta}+ \frac{\Delta U \sigma^2}{(r-r_+)^2}\right) dr^2
\nn
\\
&=& g_S + \Delta U d \tau^2 + \frac{2 \Delta U \sigma}{r -  r_+} d \tau dr + \underbrace{\frac{(r-r_+)^2+\Delta^2  \sigma^2 U}{\Delta (r-r_+)^2}}_{V} dr^2
\,.
\eena
In order to obtain a smooth metric in the domain of outer communication the constant $\sigma$ has to be chosen so that the numerator of $V$ has a triple-zero at $r=r_+$. A {\sc Mathematica}
computation gives an explicit formula for the desired constant $\sigma$, which is too lengthy to be explicitly presented here. This establishes smooth extendibility of the metric in suitable coordinates across $r=r_+$.

\subsection{Asymptotic behaviour}
 \label{s24IX17.1}

When $P=0$ the Rasheed metrics satisfy the KK-asymptotic flatness conditions. This can be seen by introducing   manifestly-asymptotically-flat coordinates $(t,x,y,z)$ in the usual way. With some work one finds that the metric takes the form
%
\bel{31III17.11}
  \left(
\begin{array}{ccccc}
 \frac{2 M}{r}+\frac{2 \Sigma }{\sqrt{3} r}-1
   & 0 & 0 & 0 & \frac{2 Q}{r} \\
 0 & \frac{2 M x^2}{r^3}-\frac{2 \Sigma
   }{\sqrt{3} r}+1 & \frac{2 M x y}{r^3} &
   \frac{2 M x z}{r^3} & 0 \\
 0 & \frac{2 M x y}{r^3} & \frac{2 M
   y^2}{r^3}-\frac{2 \Sigma }{\sqrt{3} r}+1 &
   \frac{2 M y z}{r^3} & 0 \\
 0 & \frac{2 M x z}{r^3} & \frac{2 M y z}{r^3}
   & \frac{2 M z^2}{r^3}-\frac{2 \Sigma
   }{\sqrt{3} r}+1 & 0 \\
 \frac{2 Q}{r} & 0 & 0 & 0 & \frac{4 \Sigma
   }{\sqrt{3} r}+1 \\
\end{array}
\right)
 + O(r^{-2})
 \,.
\ee
%


It turns out that when $P\ne 0$,  the Rasheed metrics do \emph{not} satisfy the KK-asymptotic flatness requirements anymore: Indeed, the phase space decomposes into sectors, labelled by $P\in \R$, in which the metrics $g$ asymptote to
the background metric
\begin{equation}
 \backg
  :=
 \left(
  dx^4+2P\cos(\theta) d\varphi
 \right)^2
  -
  dt^2
 +
  dr^2
 +
 r^2 d\theta^2
 +
 r^2 \sin^2(\theta) d\varphi^2
  \,.
 \label{11IV17.1}
\end{equation}

The metrics \eq{16VI10.1} and \eq{11IV17.1} are singular at $\sin(\theta) =0$. This can be resolved by replacing $x^4$ by $\xp^4$, respectively by $\xm^4$, on the following coordinate patches:
\bel{22IV17.5}
 \left\{
   \begin{array}{l}
    \xp^4 := x^4 + 2 P \varphi  \,,\quad \theta\in[0,\pi) \,, \\
     \xm^4 := x^4 - 2 P \varphi   \,,\quad  \theta\in(0,\pi] \,.
   \end{array}
 \right.
\ee
Indeed, the one-form
$$
 dx^4 + 2 P \cos(\theta) d\varphi = d\xp^4 + 2P (\cos(\theta) -1) d\varphi
   = d\xp^4 - \frac{2P}{r(r+z)} (x dy - y dx)
$$
is smooth for $r>0$ on  $\{\theta\in[0,\pi)\}$. Similarly the one-form
$$
 dx^4 + 2 P \cos(\theta) d\varphi = d\xm^4 + 2P (\cos(\theta) +1) d\varphi
  = d\xm^4 + \frac{2P}{r(r-z)} (x dy - y dx)
$$
is smooth on  $\{\theta\in(0,\pi]\,,\, r>0\}$. Smoothness of both $g$ and $\backg$ in the d.o.c., under the constraints discussed above,   readily follows.

We note the relation
\bel{22IV17.6}
    \xp^4 = \xm^4 + 4 P \varphi
 \,,
\ee
which implies a smooth geometry with periodic coordinates $\xp^4$ and $\xm^4$ if and only if
\bel{22IV17.7}
 \mbox{both $\xp^4$ and $\xm^4$ are periodic with period $8 P \pi$.}
\ee
From this perspective $x^4$ is \emph{not a coordinate} anymore: instead the basic coordinates are  $\xp^4$ for $\theta\in[0,\pi)$ and $\xm^4$  for $\theta\in(0,\pi]$, with $dx^4$ (but \emph{not} $x^4$) well defined away from the axes of rotation $\{\sin(\theta)=0\}$ as
\bel{22IV17.6+}
 dx^4 =
 \left\{
   \begin{array}{ll}
    d\xp^4- 2 P d\varphi  \,, & \theta\in[0,\pi) \,, \\
     d\xm^4+ 2 P d\varphi   \,, &  \theta\in(0,\pi] \,.
   \end{array}
 \right.
\ee
%

%
%

\subsubsection{Curvature of the asymptotic background}
 \label{ss26V17.1}

We continue with a calculation of the curvature tensor of the asymptotic background. It is convenient to work in the coframe
\begin{equation}
 \label{19IV17.2}
   \overline{\Theta}^\fr     0
   =
   dt
   \,,
   \qquad
   \overline{\Theta}^\fr     1
   =
   dx
   \,,
   \qquad
     \overline{\Theta}^\fr     2
   =
   dy
   \,,
   \qquad
    \overline{\Theta}^\fr     3
    =
   dz
    \,,
   \qquad
    \overline{\Theta}^\fr     4
    =
    dx^4+2P\cos(\theta) d\varphi
    \,,
\end{equation}
which is manifestly smooth after replacing $dx^4$ as in \eq{22IV17.6+}.
Using
\bel{23IV17.1}
 d  \overline{\Theta}^\fr     4 = - 2 P \sin(\theta) \, d\theta \wedge d \varphi =
- 2 P \frac{x^i}{r^3} \partial_i \rfloor (dx \wedge dy\wedge dz)= - \frac{P}{r^3} \mathring \epsilon_{\fr i \fr j \fr k} x^{\fr i} dx^{\fr j} \wedge dx^{\fr k}
 \,,
\ee
where $\mathring \epsilon_{\fr i \fr j \fr k}\in \{0,\pm 1\}$ denotes the usual epsilon {symbol}, one finds the following non-vanishing connection coefficients
\bel{25IV17.1}
 \overline{\omega}^ \fr 4{}_\fr i = \frac{ P}{r^3}  \mathring \epsilon_{ \fr i \fr  j \fr  k} x^{\fr j} \overline{\Theta}^{\fr k}
 \,,
 \quad
 \overline{\omega}^\fr i{}_\fr j
  =
    \frac{ P}{r^3} \mathring \epsilon_{\fr i \fr j \fr k} x^{\fr k} \overline{\Theta}^ \fr 4
   \,,
\ee
where $x^\fr i \equiv x^i$.
This leads to the following curvature forms
\begin{eqnarray}
\nn
 \overline{\Omega}{^\fr i {}_{\fr j}}
  &=&
  \frac{P}{r^3}
  \mathring \epsilon_{\fr i \fr j \fr k}
  \left(
   - \frac{3}{r^2} x^{\fr k}x^{\fr \ell } + \delta^{\fr k}_{\fr \ell }
  \right)
     \overline{\Theta}^{ \fr \ell } \wedge \overline{\Theta}^{ \fr 4}
  - \frac{2 P^2}{r^6} \mathring \epsilon_{\fr i \fr m  ( \fr k} \epsilon_{\fr j ) \fr n \fr \ell }
   x^{\fr m} x^{\fr n}  \overline{\Theta}^{\fr k} \wedge \overline{\Theta}^{\fr \ell }
   \,,
   \\
   \Ome{^{\fr 4}_{\fr i}}
   &=&
   \frac{P}{r^3} \mathring \epsilon_{\fr i \fr j \fr k}
     \left(
      -\frac{3}{r^2} x^{\fr j} x^{\fr \ell } + \delta^{\fr j}_{\fr \ell }
     \right)
     \overline{\Theta}^{\fr \ell } \wedge \overline{\Theta}^{\fr k}
     +
      \frac{P^2}{r^6}
      \mathring \epsilon_{\fr k \fr m \fr j} \mathring \epsilon_{\fr k \fr i \fr \ell }
      x^{\fr m} x^{\fr \ell } \overline{\Theta}^{\fr j} \wedge \overline{\Theta}^{\fr 4}
       \,,
\end{eqnarray}
hence the following non-vanishing curvature tensor components
\begin{eqnarray}
\nn
 \bKKR{^{\fr i}_{\fr j \fr k \fr 4}}
 &=&
 \frac{P}{r^3}
  \mathring \epsilon_{\fr i \fr j \fr \ell }
  \left(
   - \frac{3}{r^2} x^{\fr \ell }x^{\fr k} + \delta^{\fr \ell }_{\fr k}
  \right)
  \,,
  \\
  \bKKR{^{\fr 4}_{\fr i \fr j \fr 4}}
  &=&
 \frac{P^2}{r^6}
      \mathring \epsilon_{\fr k \fr m \fr j} \mathring \epsilon_{\fr k \fr i \fr \ell }
      x^{\fr m} x^{\fr \ell }
      \,,
 \quad
   \bKKR{^{} _{\fr i \fr j \fr k \fr \ell}} =
  - \frac{2 P^2}{r^6}(
   \mathring \epsilon_{\fr i\fr j  \fr n} \mathring \epsilon _{   \fr k \fr \ell \fr m}
   +
   \mathring \epsilon_{\fr i \fr m  [ \fr k} \mathring \epsilon_{\fr \ell] \fr j   \fr n })
   x^{\fr m} x^{\fr n}
  \,.
   \label{19V17.2}
\end{eqnarray}
The non-vanishing components of the Ricci tensor read
\begin{equation}
 \bKKR{_{\fr i \fr j}}
 =
 - \frac{2 P^2}{r^6} \mathring \epsilon_{\fr k \fr m  \fr i} \mathring \epsilon_{\fr k \fr n \fr j}
   x^{\fr m} x^{\fr n}
  \,,
  \qquad
  \bKKR{_{\fr 4 \fr 4}}
  =
   -\frac{P^2}{r^6}
      \mathring \epsilon_{\fr k \fr m \fr i} \mathring \epsilon_{\fr k \fr i \fr \ell }
      x^{\fr m} x^{\fr \ell }
  \,.
   \label{19V17.1}
\end{equation}
Subsequently the Ricci scalar is $\bKKR{}=-2P^2/r^4$.

\subsection{Global charges: a summary}

For ease of future reference we summarise the global charges of the Rasheed metrics: Let $p_\mu$ be the Hamiltonian momentum of the level sets of $t$, and let $p_{\mu,\spADM}$ be the ADM four-momentum of the space-metric $g_{ij}dx^i dx^j$. Then:
\bel{31III17.12++}
   p_{i,\spADM}
   = p_{i } = 0
  \,,
  \quad
 p_{0,\spADM} = \nored{M-\frac{\Sigma}{\sqrt{3}}}
  \,,
  \quad
  p_0 = \left\{
         \begin{array}{ll}
           2 \pi M, & \hbox{$P=0$;} \\
 4\pi  P M, & \hbox{$P\ne 0$,}
         \end{array}
       \right.
  \quad
   p_{4 } = \left\{
         \begin{array}{ll}
          2 \pi Q, & \hbox{$P=0$;} \\
  8\pi  P Q, & \hbox{$P\ne 0$.}
         \end{array}
       \right.
\ee
%
The Komar integrals associated with $X=\partial_t$ are
\bel{31III17.13}
 \frac{1}{8\pi  }
   \lim_{R\rightarrow\infty}\int_{S(R)} \int_{S^1}
    X^{\alpha;\beta} dS_{\alpha\beta}
     =
       \left\{
         \begin{array}{ll}
           2 \pi \big(M+ \frac{\Sigma}{\sqrt 3}
      \big), & \hbox{$P=0$;} \\
          8  \pi P\big(M+ \frac{\Sigma}{\sqrt 3}
      \big), & \hbox{$P\ne 0$,}
         \end{array}
       \right.
\ee
The Komar integrals associated with $X=\partial_4$ are
\bel{31III17.13+}
 \frac{1}{8\pi  }
   \lim_{R\rightarrow\infty}\int_{S(R)} \int_{S^1 }
    X^{\alpha;\beta} dS_{\alpha\beta}
     =
       \left\{
         \begin{array}{ll}
          4\pi Q, & \hbox{$P=0$;} \\
          16\pi P Q, & \hbox{$P\ne 0$.}
         \end{array}
       \right.
\ee

\section{The vector field $Z$}
 \label{A7V17.1}
Let
$$
 Z=r\partial_r
 \,.
$$
We wish to calculate  $
 \overline \nabla_\mu Z_\nu  $ for the Kottler metrics and the Rasheed metrics.

Let, first, $\backg$ be the $(n+1)$-dimensional anti-de Sitter (Kottler) metric,
\bel{24V17.21}
  \backg = -V dt^2 + V^{-1} dr^2 + r^2 \overline h
  \,,
\ee
  with
\bel{24V17.23}
 V= \lambda r^2 + \kappa
 \,,
\ee
  where $\kappa\in \{0,\pm 1\}$ is a constant,
\bel{24V17.22}
   \lambda = - \frac {2 \Lambda} {n(n-1)}
   \,,
\ee
   and where $\overline h $ is an ($r$-independent)  Einstein metric on an $(n-1)$-dimensional compact manifold $\mcK$, with scalar curvature $(n-1)(n-2)\kappa$.
    It holds that (cf., e.g., \cite{Birmingham})
    \bel{23V17.21}
     \overline R = -n (n+1) \lambda
     \,.
    \ee
    Further,
\bea
\nn
 \overline \nabla_{(\mu}Z_{\nu)} dx^\mu\otimes dx^\nu & = &
 \frac 12 \mcL_Z \backg = \frac 12 (Z^\alpha \partial_\alpha \backg_{\mu\nu}
  +\partial_\mu Z^\alpha \backg_{\alpha\nu} + \partial_\nu Z^\alpha \backg_{\alpha\mu})
  dx^\mu dx^\nu
\\
\nn
 & = &
   \frac 12 \bigg(
  r\big 
   (\partial_r(-V)dt^2 +\partial_r( V^{-1}) dr^2 +\partial_r( r^2) d\Omega
    \big) + 2    V^{-1} dr^2
   \bigg)
\\
 & = &
   \frac 12 \bigg( \frac{r \partial_r V}V ( -V dt^2) + (2  - r V^{-1} \partial_r V) V^{-1}dr^2 +    2 r^2 d\Omega^2
   \bigg)    \,,
\\
 \overline \nabla_{[\mu}Z_{\nu]} dx^\mu\otimes dx^\nu
  & = &
    \partial_{[\mu}Z_{\nu]} dx^\mu\otimes dx^\nu
    =0
 \,.
\eeal{5V17.2}
Adding, we find
\bel{5V17.1}
 \overline \nabla_\mu Z_\nu \, dx^\mu\otimes dx^\nu =   \backg \mod(\delta_\mu^t, \delta_\mu^r)
 \,,
\ee
which gives \eq{5V17.4}.

Next, for  the Rasheed background metrics \eq{11IV17.1} one finds
\bel{8V17.61}
\mcL_Z \backg = 2(dr^2 + r^2 d\Omega^2)
 \,,
 \quad
 d(\backg_{\alpha \beta}Z^\alpha dx^\beta) = d (rdr) = 0
 \,,
\ee
and \eq{5V17.4} without the $o(r^{-\gamma})$ term
 readily follows.

 \section{An identity for the Riemann tensor}
  \label{sA2III17.1}

We write $\delta^{\alpha\beta}_{\gamma\delta}$ for $ \delta^{[\alpha}_\gamma \delta^{\beta]}_\delta\equiv \frac 12 (  \delta^{ \alpha}_\gamma \delta^{\beta }_\delta-  \delta^{\beta}_\gamma \delta^{\alpha}_\delta)$, etc.

For completeness we prove the following identity satisfied by the Riemann tensor,
 which is valid in any dimension,  is clear in dimensions two and three, implies the double-dual identity for the Weyl tensor in dimension four,
 and is probably well known in higher dimensions as well:
\begin{equation}
 \label{2III17.3}
  \delta^{\alpha\beta\gamma\delta}_{\mu\nu\rho\sigma} \tensor{R}{^\rho^\sigma_\gamma_\delta}
  =
  \frac{1}{3!}
   \left(
    \tensor{R}{^\alpha^\beta_\mu_\nu}+\delta^{\alpha\beta}_{\mu\nu}
    R-4\delta^{[\alpha}_{[\mu} \tensor{R}{^{\beta]}_{\nu]}}
  \right)
  \,.
\end{equation}
The above holds for any tensor field satisfying
\bel{25VI17.2}
 R_{\alpha\beta\gamma\delta} =
- R_{\beta\alpha\gamma\delta} =
  R_{\beta\alpha\delta\gamma}
 \,.
\ee
To prove \eq{2III17.3} one can calculate as follows:
%
\begin{eqnarray}
 \nn
 4! \,\delta^{\alpha\beta\gamma\delta}_{\mu\nu\rho\sigma} \tensor{R}{^\rho^\sigma_\gamma_\delta}
  &=&
   2\big[\delta^\alpha_\mu
    \left(
     \delta^\beta_\nu \delta^\gamma_\rho
     \delta^\delta_\sigma-\delta^\beta_\rho \delta^\gamma_\nu
     \delta^\delta_\sigma+\delta^\beta_\sigma \delta^\gamma_\nu
     \delta^\delta_\rho
    \right)
    \\
    \nn
  &&
   -
   \delta^\alpha_\nu
   \left(
    \delta^\beta_\mu \delta^\gamma_\rho
    \delta^\delta_\sigma-\delta^\beta_\rho \delta^\gamma_\mu
    \delta^\delta_\sigma+\delta^\beta_\sigma \delta^\gamma_\mu
    \delta^\delta_\rho
   \right)
   \\
   \nn
  &&
   +
   \delta^\alpha_\rho
   \left(
   \delta^\beta_\mu \delta^\gamma_\nu
   \delta^\delta_\sigma-\delta^\beta_\nu \delta^\gamma_\mu
   \delta^\delta_\sigma+\delta^\beta_\sigma \delta^\gamma_\mu
   \delta^\delta_\nu \right)
   \\
   \nn
  &&
   -
   \delta^\alpha_\sigma
   \left(\delta^\beta_\mu \delta^\gamma_\nu \delta^\delta_\rho-\delta^\beta_\nu
   \delta^\gamma_\mu \delta^\delta_\rho+\delta^\beta_\rho \delta^\gamma_\mu
   \delta^\delta_\nu\right)\big]\tensor{R}{^\rho^\sigma_\gamma_\delta}
   \\
   \nn
  &=&
  2
  \left(
    2\delta^{\alpha\beta}_{\mu\nu} \delta^\gamma_\rho\delta^\delta_\sigma
    -
    4\delta^{\alpha\gamma}_{\mu\nu}\delta^{\beta\delta}_{\rho\sigma}
    +
    4\delta^{\beta\gamma}_{\mu\nu} \delta^{\alpha\delta}_{\rho\sigma}
    +
    2\delta^\alpha_\rho
    \delta^\beta_\sigma \delta^\gamma_\mu \delta^\delta_\nu \right)\tensor{R}{^\rho^\sigma_\gamma_\delta}
   \\
   \nn
 &=&
  4\left(
   \delta^{\alpha\beta}_{\mu\nu}  R^{\gamma\delta}{}_{\gamma \delta}
   -
   2\delta^{\alpha\gamma}_{\mu\nu} \tensor{R}{^\beta^\sigma_\gamma_\sigma}
   +
   2\delta^{\beta\gamma}_{\mu\nu} \tensor{R}{^\alpha^\sigma_\gamma_\sigma}
   +
   \tensor{R}{^\alpha^\beta_\mu_\nu}
  \right)
  \\
 &=&
  4\left(
     \tensor{R}{^\alpha^\beta_\mu_\nu}
     +
     \delta^{\alpha\beta}_{\mu\nu}R^{\gamma\delta}{}_{\gamma \delta}
     -
     4\delta^{[\alpha}_{[\mu} \tensor{R}{^{\beta]\gamma}_{\nu]\gamma }}
    \right)
  \,.
  \label{2III17.4}
\end{eqnarray}
If the sums are over all  indices we obtain \eq{2III17.3}. The reader is warned, however, that in some of our calculations the sums will be only over a subset of all possible indices, in which case
the last equation remains valid but the last two terms in \eq{2III17.4} \emph{cannot} be replaced by the Ricci scalar and the Ricci tensor.

\medskip

\noindent{\sc Acknowledgements:} Useful discussions with Abhay Ashtekar, and comments from Eric Woolgar  are acknowledged.

\end{document}